\documentclass[11pt,letterpaper]{article}
\usepackage[T1]{fontenc}
\usepackage[utf8]{inputenc}
\usepackage[margin=1in]{geometry}
\usepackage{newtxtext}
\usepackage{amsmath,amsfonts,amssymb,amsthm}
\usepackage[subscriptcorrection]{newtxmath}
\usepackage{graphicx,placeins,float,booktabs,longtable,xcolor}
\usepackage{microtype}
\usepackage[round,authoryear]{natbib}
\usepackage[font=small,labelfont=bf]{caption}
\usepackage[hidelinks,unicode,bookmarksnumbered]{hyperref}
\usepackage{bookmark}
\usepackage[nameinlink,noabbrev]{cleveref}
\graphicspath{{fig/}}
\hypersetup{
  pdftitle={Score-aware nonparametric adaptation of prediction-powered inference},
  pdfauthor={Xin Lu, Xuming He, Bo Li},
  pdfsubject={Prediction-powered inference},
  pdfkeywords={Dimension reduction, Generalized linear models, Prediction-powered inference, Spline approximation}
}

\newtheorem{theorem}{Theorem}
\newtheorem{lemma}{Lemma}
\newtheorem{corollary}{Corollary}
\newtheorem{proposition}{Proposition}
\theoremstyle{definition}
\newtheorem{assumption}{Assumption}
\newtheorem{remark}{Remark}
\newenvironment{keywords}{\par\smallskip\noindent\textbf{Keywords:}\ }{\par\medskip}
\crefname{assumption}{Assumption}{Assumptions}
\Crefname{assumption}{Assumption}{Assumptions}
\crefname{theorem}{Theorem}{Theorems}
\Crefname{theorem}{Theorem}{Theorems}
\crefname{lemma}{Lemma}{Lemmas}
\Crefname{lemma}{Lemma}{Lemmas}
\crefname{proposition}{Proposition}{Propositions}
\Crefname{proposition}{Proposition}{Propositions}
\crefname{corollary}{Corollary}{Corollaries}
\Crefname{corollary}{Corollary}{Corollaries}
\crefname{remark}{Remark}{Remarks}
\Crefname{remark}{Remark}{Remarks}

\newenvironment{assumptionparts}
{\leavevmode\par\nobreak
 \begin{list}{\textup{(\roman{enumi})}}{
    \usecounter{enumi}
    \setlength{\leftmargin}{2.5em}
    \setlength{\labelwidth}{2em}
    \setlength{\labelsep}{0.5em}
    \setlength{\itemindent}{0pt}
    \setlength{\listparindent}{0pt}
    \setlength{\topsep}{0.25em}
    \setlength{\itemsep}{0.25em}
    \setlength{\parsep}{0pt}}}
{\end{list}}

\DeclareMathOperator*{\argmin}{arg\,min}

\DeclareMathOperator{\E}{E}
\DeclareMathOperator{\Var}{var}
\newcommand{\bs}{\boldsymbol}

\newcommand{\ppi}{\mathrm{ppi}}

\DeclareMathOperator{\Cov}{cov}
\newcommand{\tr}{\operatorname{tr}}

\newcommand{\PPI}{\texttt{PPI}}

\newcommand{\thetalab}{\widehat{\bs\theta}_{\rm lab}}

\title{Score-aware nonparametric adaptation of prediction-powered inference}
\author{Xin Lu \qquad Xuming He \qquad Bo Li\thanks{Corresponding author. Email: \href{mailto:bol@wustl.edu}{bol@wustl.edu}.}\\[0.5em]
\small Department of Statistics and Data Science\\
\small Washington University in St. Louis}
\date{}

\begin{document}
\maketitle

\begin{abstract}
Prediction-powered inference supplements labelled data with unlabelled data and external predictions to improve statistical efficiency. For inference on generalized linear model (GLM) coefficients, optimal use of these predictions generally requires estimating a conditional mean given the covariates and the external prediction, which can be challenging when the labelled data are limited or the covariates are multi-dimensional.
We address this challenge through dimension reduction. We first fit an augmented generalized linear model combining the covariates and external prediction to obtain a scalar index, and then approximate the conditional mean by a univariate function of this index. Based on this construction, we develop a class of estimators for GLM coefficients and select the proposed estimator by minimizing the estimated asymptotic covariance trace.
Under regularity conditions, the proposed estimator is asymptotically normal and has an asymptotic covariance trace no greater than that of the labelled-only estimator, even when the dimension reduction is insufficient. We also establish conditions under which the dimension reduction is sufficient. Simulations show that the proposed estimator is generally close to the best feasible estimator in root mean squared error, with confidence interval coverage close to the nominal level.

\end{abstract}


\section{Introduction}

Externally trained models can provide predictions for unobserved outcomes, creating opportunities to use large unlabelled samples to improve the efficiency of estimating population quantities \citep{mccaw2024synthetic}. However, treating these predictions as observed outcomes can introduce bias. Prediction-powered inference (\PPI{}) addresses this issue by using labelled observations to correct the resulting bias, thereby allowing external predictions to contribute to statistical inference without requiring them to be accurate \citep{angelopoulos2023prediction}.

Although \PPI{} provides valid inference, it can be less efficient than labelled-only inference when the predictions are insufficiently informative. Subsequent methods have therefore been developed to adapt the contribution of the predictions to guard against such efficiency losses  \citep{angelopoulos2023ppi++,miao2025assumption,gan2024prediction,gronsbell2024another}. More recently, \citet{ji2025predictions} characterized the optimal use of predictions in \PPI{} and proposed recalibrated \PPI{}. Their approach requires estimating, from the labelled sample, a multivariate conditional mean given the covariates and external prediction. This estimation can be difficult when the labelled sample is limited, or when the covariate dimension is moderately high. We therefore develop a score-aware nonparametric (SAN) estimator based on dimension reduction to mitigate the resulting curse of dimensionality in the conditional mean estimation.

We focus on inference for the coefficients of a generalized linear model (GLM). Our contributions are threefold. First, we propose a dimension reduction approach to the conditional mean estimation required for optimal \PPI{}. We fit an augmented GLM combining the covariates and external prediction to construct a scalar index, and then approximate the conditional mean by a univariate function of this index.
Second, to guard against efficiency loss from a possibly poor conditional mean approximation, we use spline functions of the scalar index to construct a class of predicted scores that includes the labelled-only estimator as a special case. We then select the proposed estimator by minimizing the estimated asymptotic covariance trace over this class. The proposed estimator has an asymptotic covariance trace no greater than that of the labelled-only estimator, even when the dimension reduction is insufficient.
Third, we establish asymptotic normality and consistent variance estimation under regularity conditions.  We further give conditions under which the scalar index is sufficient for the conditional mean and hence preserves the information needed for optimal \PPI{} \citep{li1989regression}.

Simulations show that, when the conditional mean has a low-dimensional structure, the proposed estimator is generally close to the best feasible estimator in terms of root mean squared error, while its confidence intervals achieve empirical coverage close to the nominal level.

Throughout, \(\|\cdot\|\) denotes the Euclidean norm of a vector,
and \(\|\cdot\|_F\) the Frobenius norm of a matrix.
For a matrix \(\bs A\), \(\bs A^\top\) denotes its transpose, and when \(\bs A\) is invertible, \(\bs A^{-\top}=(\bs A^{-1})^\top\).
We write \(\tr(\bs A)\) for the trace of $\bs A$ and
\(\lambda_{\min}(\bs A)\) for the smallest eigenvalue of $\bs A$.
For symmetric matrices \(\bs A\) and \(\bs B\) of the same dimension,
\(\bs A\succeq\bs B\) means that \(\bs A-\bs B\) is positive semidefinite.
The symbols \(\xrightarrow{\mathbb P}\) and \(\rightsquigarrow\)
denote convergence in probability and in distribution, respectively.

\section{Prediction-powered inference for generalized linear models}
\label{sec:glm-ppi}

\subsection{Data and target parameter}

Consider inference for a generalized linear model (GLM)
with a canonical link \citep{mccullagh1989generalized}.
Let \(Y\) denote the outcome, and let
\(\widetilde{\bs X}=(1,\bs X^\top)^\top\in\mathbb R^p\)
denote the covariate vector including an intercept,
where \(\bs X\in\mathbb R^{p-1}\) contains the random
covariates.
The working GLM specifies the conditional density
\begin{equation}
    p_{\bs\theta,\phi}(y\mid\bs X)
    =
    \exp\left[{\frac 1 \phi}
        \left\{
            y\widetilde{\bs X}^{\top}\bs\theta
            -A(\widetilde{\bs X}^{\top}\bs\theta)
        \right\}
        +c(y,\phi)
    \right],
    \label{eq:glm-density}
\end{equation}
where \(\bs\theta\in\mathbb R^p\)
is the parameter of interest,
\(\phi>0\) is the dispersion parameter,
and \(A\) and \(c\) are known functions.
Let \(\Theta\subseteq\mathbb R^p\) be a convex parameter space.
The target parameter \(\bs\theta^\ast\) is the unique solution in the interior of \(\Theta\) to the population score equation
\begin{equation}
    \E\{\bs\psi_{\bs\theta}(Y,\bs X)\}=\bs0,
    \label{eq:glm-score}
\end{equation}
where
$\bs\psi_{\bs\theta}(Y,\bs X)= \widetilde{\bs X}\{Y-\mu_{\bs\theta}\}$,
$\mu_{\bs\theta}=A'(\widetilde{\bs X}^{\top}\bs\theta)$,
and \(A'\) denotes the derivative of \(A\).
The working GLM is used to define
\(\bs\theta^\ast\), i.e., the conditional distribution of \(Y\)
given \(\bs X\) need not follow this model.
Under misspecification, \(\bs\theta^\ast\) is the
corresponding pseudo-true parameter
\citep{white1982maximum,mccullagh1989generalized}.

To infer $\bs\theta^\ast$, we consider a prediction-powered setting \citep{angelopoulos2023prediction} with a labelled sample of size $n$ and an unlabelled sample of size $N$. The covariates $\bs X$ are observed in both samples, whereas the outcome $Y$ is observed only in the labelled sample.
An externally trained prediction model provides a scalar prediction $f$ of $Y$ for each unit in both samples. The model may use high-dimensional or unstructured auxiliary information beyond $\bs X$. Accordingly, the observed data consist of $(Y,f,\bs X)$ in the labelled sample and $(f,\bs X)$ in the unlabelled sample. The following assumption formalizes the sampling framework for prediction-powered inference.

\begin{assumption}
\label{ass:sampling}
The labelled observations \(\{(Y_i,f_i,\bs X_i)\}_{i=1}^n\)
are i.i.d. copies of \((Y,f,\bs X)\), and the unlabelled observations
\(\{(f_i,\bs X_i)\}_{i=n+1}^{n+N}\)
are i.i.d. copies of \((f,\bs X)\).
The two samples are independent.
\end{assumption}

For any quantity observed in the relevant sample, write
\(\mathbb E_n a=n^{-1}\sum_{i=1}^n a_i\)
and
\(\mathbb E_N a=N^{-1}\sum_{i=n+1}^{n+N}a_i\).
As a benchmark and a starting point for the prediction-powered estimators developed below, we consider the labelled-only estimator $\widehat{\bs\theta}_{\rm lab}$, which solves
$\mathbb E_n\bs\psi_{\bs\theta}(Y,\bs X)=\bs0$.
Under standard regularity conditions, as \(n\to\infty\),
we have \citep{newey1994large}:
\[
    \sqrt n(\widehat{\bs\theta}_{\rm lab}-\bs\theta^\ast)
    \rightsquigarrow \mathcal{N}(\bs0,\bs\Sigma_{\rm lab}),
\]
where, with \(\dot{\bs\psi}_{\bs\theta}\) denoting the derivative of \(\bs\psi_{\bs\theta}\) with respect to \(\bs\theta\), and \(A''\) denoting the second derivative of \(A\),
\[
\bs\Sigma_{\rm lab}
    =\bs J^{-1}
    \Cov\{\bs\psi_{\bs\theta^\ast}(Y,\bs X)\}
    \bs J^{-\top},
    \qquad
    \bs J
    =-\E\{\dot{\bs\psi}_{\bs\theta^\ast}(Y,\bs X)\}
    =\E\left[
        A''(\widetilde{\bs X}^{\top}\bs\theta^\ast)
        \widetilde{\bs X}\widetilde{\bs X}^{\top}
    \right].
\]

Before introducing our proposed estimator, we briefly review the relevant \PPI{} methodology in Section \ref{sec:PPI-review} to motivate our approach.

\subsection{Prediction-powered inference and optimal predicted scores}\label{sec:PPI-review}

To utilize the unlabelled data, PPI methods predict the score
\(\bs\psi_{\bs\theta}(Y,\bs X)\)
for the unlabelled units using a given family of functions
\(\{\bs h_{\bs\theta}(f,\bs X):\bs\theta\in\Theta\}\).
The labelled sample is then used to correct the bias
introduced by this prediction.
The corresponding PPI estimator, denoted by
\(\widehat{\bs\theta}_{\bs h}\), solves
\begin{equation}
    \label{eq:ppi-score}
    \mathbb E_N\bs h_{\bs\theta}(f,\bs X)
    +
    \mathbb E_n\{
        \bs\psi_{\bs\theta}(Y,\bs X)
        -
        \bs h_{\bs\theta}(f,\bs X)
    \}
    =\bs0.
\end{equation}
Under Assumption~\ref{ass:sampling}, the bias correction ensures
that the left-hand side of \eqref{eq:ppi-score} has expectation
\(\E\{\bs\psi_{\bs\theta}(Y,\bs X)\}\).
The original PPI method, denoted by
\(\widehat{\bs\theta}_{\ppi}\), uses
\(\bs h_{\bs\theta}(f,\bs X)\equiv\bs\psi_{\bs\theta}(f,\bs X)\),
replacing \(Y\) with \(f\) in the score function \citep{angelopoulos2023prediction}.

Under some regularity conditions, if $n,N \to  \infty$ and $n/(N+n) \to \rho \in [0,1)$, \(\widehat{\bs\theta}_{\bs h}\)
is asymptotically normal provided that \(\widehat{\bs\theta}_{\bs h}\) is consistent
\citep{gan2024prediction,ji2025predictions}:
\begin{align}
      \label{eq:ppi-h}
      \sqrt n(\widehat{\bs\theta}_{\bs h}-\bs\theta^\ast)
    \rightsquigarrow \mathcal{N}\{\bs0,\bs\Sigma(\bs h)\},
\end{align}
where
\begin{align*}
    \bs\Sigma(\bs h)
    &=\bs J^{-1}\bs\Omega(\bs h)\bs J^{-\top},\\
    \bs\Omega(\bs h)
    &=\Cov\{
        \bs\psi_{\bs\theta^\ast}(Y,\bs X)
        -\bs h_{\bs\theta^\ast}(f,\bs X)
    \}
    +\frac{\rho}{1-\rho}
    \Cov\{\bs h_{\bs\theta^\ast}(f,\bs X)\}.
\end{align*}

Since \eqref{eq:ppi-score} may not have a unique root,
an alternative is the following one-step estimator,
which has the same asymptotic distribution as in
\eqref{eq:ppi-h} under the same regularity conditions:
\begin{equation}
\begin{aligned}
    \widetilde{\bs\theta}_{\bs h}
    &=\widehat{\bs\theta}_{\rm lab}
    +\widehat{\bs J}^{-1}
    (\mathbb E_N-\mathbb E_n)
    \bs h_{\widehat{\bs\theta}_{\rm lab}}(f,\bs X),
\qquad \widehat{\bs J}
    &=\mathbb E_n\left[
        A''(\widetilde{\bs X}^{\top}\widehat{\bs\theta}_{\rm lab})
        \widetilde{\bs X}\widetilde{\bs X}^{\top}
    \right].
\end{aligned}
\label{eq:one-step-ppi}
\end{equation}

The asymptotic covariance \(\bs\Sigma(\bs h)\) depends on the predicted-score
family $\bs h_{\bs\theta}(f,\bs X)$ only through its value at \(\bs\theta^\ast\).
We therefore write \(\bs h(f,\bs X)=\bs h_{\bs\theta^\ast}(f,\bs X)\).
An optimal choice of $\bs h$ minimizing \(\bs\Sigma(\bs h)\) \citep{ji2025predictions} is
\begin{equation}
\begin{aligned}
    \bs h^{\mathrm{opt}}(f,\bs X)
    &=
    (1-\rho)
    \E\{\bs\psi_{\bs\theta^\ast}(Y,\bs X)\mid f,\bs X\}\\
    &=
    (1-\rho)\widetilde{\bs X}
    \{\E(Y\mid f,\bs X)-\mu_{\bs\theta^\ast}\}.
\end{aligned}
\label{eq:h-full}
\end{equation}

Apart from the parameter $\bs\theta^\ast$, the optimal predicted score in \eqref{eq:h-full}  depends on the unknown conditional mean $\E(Y\mid f,\bs X)$, which must be estimated from the labelled sample.
Although $f$ is scalar, estimating $\E(Y\mid f,\bs X)$ remains a regression problem with $p$ predictors. If we impose no restriction on this conditional mean, estimation can be difficult when labelled data are limited or the covariate dimension is moderately high.
This motivates the score-aware nonparametric (SAN) estimator developed in the next section, which reduces the conditional mean estimation problem to one dimension.

\section{Score-aware nonparametric adaptation}
\label{sec:dimension-reduction}

\subsection{An augmented GLM for dimension reduction and predicted-score construction}

To simplify estimation of $\E(Y\mid f,\bs X)$ in \eqref{eq:h-full}, we use an augmented GLM to combine $f$ and $\bs X$ into a scalar index.
To place the external prediction on a scale suitable for the augmented GLM without loss of information,
let \(\tau\) be a prespecified measurable one-to-one
transformation of \(f\).
When $(A')^{-1}$ is well defined on the range of $f$, we use the default transformation $\tau(f)=(A')^{-1}(f)$. Otherwise, the identity transformation $\tau(f)=f$ may be used. For example, under a logistic model with binary predictions $f\in\{0,1\}$, we use the identity transformation because the logit is not defined at 0 or 1.
Let \(\bs Q=\{\tau(f),\bs X^\top\}^\top\) and
\(\widetilde{\bs Q}=(1,\bs Q^\top)^\top\).
Let \(\bs\gamma^\ast\in\mathbb R^{p+1}\) be the unique
solution to the population score equation for the augmented GLM,
\begin{equation}
    \E\left[
        \widetilde{\bs Q}
        \{Y-A'(\widetilde{\bs Q}^{\top}\bs\gamma)\}
    \right]
    =\bs0.
    \label{eq:augmented-score}
\end{equation}
The corresponding mean function defines the scalar index
\begin{equation}
    S=A'(\widetilde{\bs Q}^{\top}\bs\gamma^\ast).
    \label{eq:population-index}
\end{equation}

We then approximate $\E(Y\mid f,\bs X)$ by a univariate spline function of $S$ with B-spline basis
$\bs b_K(S)=\{b_{1,K}(S),\ldots,b_{K,K}(S)\}^\top$,
where the basis dimension \(K=K_n\) may be fixed or increase with
the labelled sample size \(n\).
Substituting this approximation into \eqref{eq:h-full} yields predicted scores of the form
\[
    \widetilde{\bs X}
    \{\bs b_K(S)^\top\bs\beta_1
      -(1-\rho)\mu_{\bs\theta^\ast}\},
    \qquad \bs\beta_1\in\mathbb R^K,
\]
where the factor \(1-\rho\) is absorbed into
\(\bs\beta_1\).

When the approximation of $\E(Y\mid f,\bs X)$ is poor, the resulting PPI estimator may be less efficient than $\widehat{\bs\theta}_{\rm lab}$.
We therefore append
\(\mu_{\bs\theta^\ast}\) in the basis vector.
Specifically, define the basis matrix
\begin{equation}
    \bs B_{\bs\theta,K}
    =
    \widetilde{\bs X}
    [\mu_{\bs\theta},\ \bs b_K(S)^\top],
    \label{eq:B-definition}
\end{equation}
and the corresponding predicted-score class used to construct the SAN estimator,
\[
    \mathcal H_{\mathrm{san},K}
    =
    \left\{
        \bs B_{\bs\theta^\ast,K}\bs\beta:
        \bs\beta\in\mathbb R^{K+1}
    \right\}.
\]
This class includes $\widehat{\bs\theta}_{\rm lab}$ as a special case when $\bs\beta=\bs 0$, allowing us to guard against efficiency loss by optimizing some form of efficiency over $\bs \beta$.

Within \(\mathcal H_{\mathrm{san},K}\), define the oracle
coefficients for the SAN estimator by minimizing the trace of the PPI estimator's
asymptotic covariance matrix:
\begin{equation}
    \bs\beta_{\mathrm{san},K}
    =
    \argmin_{\bs\beta\in\mathbb R^{K+1}}
    \tr\left\{
        \bs\Sigma(\bs B_{\bs\theta^\ast,K}\bs\beta)
    \right\},
    \label{eq:beta-opt}
\end{equation}
where \(\bs\Sigma(\bs h)\) is defined in \eqref{eq:ppi-h}.

The criterion in \eqref{eq:beta-opt} is quadratic
in \(\bs\beta\).
To obtain its minimizer, define the population centering
operator by \(\mathcal C a=a-\E a\),
where \(a\) may be a scalar, vector or matrix.
Let \(\bs M=\bs J^{-\top}\bs J^{-1}\),
and
\begin{equation}
    \bs G_K
    =
    \E\left[
        (\mathcal C\bs B_{\bs\theta^\ast,K})^\top
        \bs M\mathcal C\bs B_{\bs\theta^\ast,K}
    \right],\quad
    \bs d_K
    =
    \E\left[
        (\mathcal C\bs B_{\bs\theta^\ast,K})^\top
        \bs M\mathcal C\bs\psi_{\bs\theta^\ast}
    \right].
\label{eq:G-d}
\end{equation}
\Cref{prop:beta-closed-form} gives the minimizer of
\eqref{eq:beta-opt} and its trace reduction relative
to labelled-only inference.

\begin{proposition}
\label{prop:beta-closed-form}
Suppose that \(\bs\psi_{\bs\theta^\ast}\) and the
columns of \(\bs B_{\bs\theta^\ast,K}\) have finite
second moments, and that \(\bs J\) and \(\bs G_K\)
are nonsingular.
Then the unique solution to \eqref{eq:beta-opt} and its trace reduction are
\[
\begin{gathered}
    \bs\beta_{\mathrm{san},K}
    =
    (1-\rho)\bs G_K^{-1}\bs d_K,\\
    \tr(\bs\Sigma_{\rm lab})
    -
    \tr\{
        \bs\Sigma(
            \bs B_{\bs\theta^\ast,K}
            \bs\beta_{\mathrm{san},K}
        )
    \}
    =
    (1-\rho)\bs d_K^\top\bs G_K^{-1}\bs d_K
    \geq0.
\end{gathered}
\]
\end{proposition}

The oracle choice therefore has an asymptotic covariance
trace no greater than that of the labelled-only estimator,
even when the conditional mean approximation is poor.
The reduction is strictly positive if and only if
\(\bs d_K\ne\bs0\). As indicated by the factor $1-\rho$, the trace reduction becomes more pronounced as the relative
size of the unlabelled sample increases.

\begin{remark}
We may not use all the columns in $ \bs B_{\bs\theta,K}$ since some may be redundant.
For example, in linear regression with only an intercept,
\(\bs B_{\bs\theta^\ast,K}
    \equiv
    [\E Y, \bs b_K(S)^\top]\).
Since the full B-spline basis satisfies
\(\sum_{k=1}^K b_{k,K}(S)=1\),
the first column of \(\mathcal C\bs B_{\bs\theta^\ast,K}\)
is zero, and the remaining columns also sum to zero. We may therefore omit the $\E Y$ column and one
B-spline column.
When a subset of the columns is used to ensure that $\bs G_K$ is non-singular,
we retain the same notation for the basis matrix
and its coefficient vector.
With a slight abuse of notation, we continue to denote
the number of columns by \(K+1\).
\end{remark}

\subsection{Estimation and inference}
We now construct the proposed SAN estimator by substituting
an estimate of
\(\bs B_{\bs\theta^\ast,K}\bs\beta_{\mathrm{san},K}\)
into the one-step estimator \eqref{eq:one-step-ppi}.
We first estimate the index $S$ from the labelled sample:
\begin{equation}
    \widehat{\bs\gamma}
    =
    \argmin_{\bs\gamma}
    \mathbb E_n\left[
        A(\widetilde{\bs Q}^{\top}\bs\gamma)
        -Y\widetilde{\bs Q}^{\top}\bs\gamma
    \right].
    \label{eq:augmented-fit}
\end{equation}
For \(i=1,\ldots,n+N\), the estimated indices and sample
counterparts of \eqref{eq:B-definition} are
\begin{align}
    \widehat S_i
    &=
    A'(\widetilde{\bs Q}_i^\top\widehat{\bs\gamma}),
    \label{eq:estimated-index}\\
    \widehat{\bs B}_{K,i}
    &=
    \widetilde{\bs X}_i
    [\widehat\mu_i,\ \bs b_K(\widehat S_i)^\top],
    \label{eq:sample-B-definition}
\end{align}
where
\(\widehat\mu_i
=A'(\widetilde{\bs X}_i^\top
\widehat{\bs\theta}_{\rm lab})\).

For labelled units, let
\(\widehat{\bs\psi}_i
=\bs\psi_{\widehat{\bs\theta}_{\rm lab}}(Y_i,\bs X_i)\),
\(i=1,\ldots,n\).
Using \(\widehat{\bs J}\) from \eqref{eq:one-step-ppi},
define
\(\widehat{\bs M}
=\widehat{\bs J}^{-\top}\widehat{\bs J}^{-1}\).
For empirical centering, write
\(\mathcal C_m A=A-\mathbb E_m A\),
\(m\in\{n,N\}\).
We have labelled-sample counterparts of \eqref{eq:G-d}:
\begin{equation}
\begin{aligned}
    \widehat{\bs G}_K
    &=
    \mathbb E_n\left[
        (\mathcal C_n\widehat{\bs B}_K)^\top
        \widehat{\bs M}
        \mathcal C_n\widehat{\bs B}_K
    \right],\quad
    \widehat{\bs d}_K
    =
    \mathbb E_n\left[
        (\mathcal C_n\widehat{\bs B}_K)^\top
        \widehat{\bs M}
        \mathcal C_n\widehat{\bs\psi}
    \right].
\end{aligned}
\label{eq:sample-G-d}
\end{equation}
The estimator of the oracle coefficients ${\bs\beta}_{\mathrm{san},K}$, denoted by \(\widehat{\bs\beta}_{\mathrm{san},K}\), is a solution to
\begin{equation}
    \widehat{\bs G}_K
    \widehat{\bs\beta}_{\mathrm{san},K}
    =
    (1-\rho_n)\widehat{\bs d}_K,
    \qquad
    \rho_n=\frac{n}{n+N}.
    \label{eq:beta-hat}
\end{equation}
 Then, our proposed SAN estimator is
\begin{equation}
    \widehat{\bs\theta}_{\mathrm{san},K}
    =
    \widehat{\bs\theta}_{\rm lab}
    +
    \widehat{\bs J}^{-1}
    \bigl\{
        (\mathbb E_N-\mathbb E_n)\widehat{\bs B}_K
    \bigr\}
    \widehat{\bs\beta}_{\mathrm{san},K}.
    \label{eq:SAN-estimator}
\end{equation}
For variance estimation, write
\(\widehat{\bs h}_{\mathrm{san},K,i}
=\widehat{\bs B}_{K,i}
 \widehat{\bs\beta}_{\mathrm{san},K}\)
for \(i=1,\ldots,n+N\), and let
\[
    \widehat{\bs V}_m(a)
    =
    \mathbb E_m\left[
        (\mathcal C_m a)(\mathcal C_m a)^\top
    \right],
    \qquad m\in\{n,N\}
\]
denote the empirical covariance matrix of vector \(a\).
Let $    d_{\rm eff}
    =
    p+(p+1)+(K+1)
    =
    2p+K+2$
count the coefficients of the GLM in
\eqref{eq:glm-score}, the augmented GLM in
\eqref{eq:augmented-fit}, and the predicted score in \eqref{eq:beta-opt}, respectively. Following the expression for \(\bs\Sigma(\bs h)\)
in \eqref{eq:ppi-h}, we estimate the asymptotic covariance
of \(\sqrt n(\widehat{\bs\theta}_{\mathrm{san},K}
-\bs\theta^\ast)\) by
\begin{equation}
\begin{aligned}
    \widehat{\bs\Sigma}_{\mathrm{san},K}
    &=
    \frac{n}{n-d_{\rm eff}}
    \widehat{\bs J}^{-1}
    \widehat{\bs\Omega}_{\mathrm{san},K}
    \widehat{\bs J}^{-\top}, \quad
    \widehat{\bs\Omega}_{\mathrm{san},K}
    &=
    \widehat{\bs V}_n
    (\widehat{\bs\psi}
      -\widehat{\bs h}_{\mathrm{san},K})
    +
    \frac{n}{N}
    \widehat{\bs V}_N
    (\widehat{\bs h}_{\mathrm{san},K}).
\end{aligned}
\label{eq:variance-estimator}
\end{equation}
The multiplier \(n/(n-d_{\rm eff})\), used when
\(d_{\rm eff}<n\), is a heuristic finite-sample adjustment.

For a fixed nonzero \(\bs c \in\mathbb R^p\),
a \(1-\alpha\) Wald confidence interval for
\(\bs c^\top\bs\theta^\ast\) is
\[
    \bs c^\top\widehat{\bs\theta}_{\mathrm{san},K}
    \ \pm\
    z_{1-\alpha/2}
    \left\{
        {
            \bs c^\top
            \widehat{\bs\Sigma}_{\mathrm{san},K}
            \bs c
        }/{n}
    \right\}^{1/2},
\]
where \(z_{1-\alpha/2}\) is the
\((1-\alpha/2)\)-quantile of the standard normal distribution.

\section{Theoretical properties}
\label{sec:theory}
\subsection{Asymptotic normality and inference validity}

We establish asymptotic normality and inference validity
for $\widehat{\bs\theta}_{\mathrm{san},K}$. Throughout this section, \(p\) is fixed.
Write \(\bs\psi^\ast=\bs\psi_{\bs\theta^\ast}(Y,\bs X)\)
and \(S_{\bs\gamma}=A'(\widetilde{\bs Q}^{\top}\bs\gamma)\).
We first impose regularity conditions on the GLM models
and the spline basis.

\begin{assumption}
\label{ass:score-learning}
\begin{assumptionparts}

\item
The matrix \(\bs J\) is nonsingular, and
\(\widehat{\bs\gamma}-\bs\gamma^\ast
=O_{\mathbb P}(n^{-1/2})\).

\item
There exist fixed open neighborhoods
\(\mathcal U_\theta\subseteq\Theta\) and
\(\mathcal U_\gamma\) of \(\bs\theta^\ast\) and
\(\bs\gamma^\ast\), respectively, on which
\(\mu_{\bs\theta}\) and \(S_{\bs\gamma}\) are twice
continuously differentiable in their respective parameters
almost surely.
There exists a measurable random variable \(D\geq1\)
with
\(\E[\|\widetilde{\bs X}\|^4(|Y|^4+D^4)]<\infty\)
such that, almost surely,
\[
\sup_{\bs\theta\in\mathcal U_\theta}
\bigl\{
    |\mu_{\bs\theta}|
    +\|\partial_{\bs\theta}\mu_{\bs\theta}\|
    +\|\partial_{\bs\theta}^2\mu_{\bs\theta}\|_F
\bigr\}
\leq D,\quad
\sup_{\bs\gamma\in\mathcal U_\gamma}
\bigl\{
    \|\partial_{\bs\gamma}S_{\bs\gamma}\|^2
    +\|\partial_{\bs\gamma}^2S_{\bs\gamma}\|_F
\bigr\}
\leq D.
\]

\item
There exists a fixed interval \([s_-,s_+]\) such that
\(S_{\bs\gamma}\in[s_-,s_+]\) for all
\(\bs\gamma\in\mathcal U_\gamma\), almost surely.
The vector \(\bs b_K\) is a standard B-spline basis of fixed
degree \(r\geq2\) on \([s_-,s_+]\), with deterministic and
quasi-uniform knots, as specified in
Supplementary Material.

\item
The matrix \(\bs G_K\) satisfies
\(\lambda_{\min}(\bs G_K)\geq c_G/K\)
for some constant \(c_G>0\) independent of \(K\).

\item
The basis dimension satisfies \(K^5/n\to0\).

\end{assumptionparts}
\end{assumption}

Condition (i) requires root-\(n\) consistency of the
augmented-GLM estimator.
Condition (ii) imposes local smoothness and moment
conditions that permit Taylor expansions in
\(\bs\theta\) and \(\bs\gamma\) and control their remainders.
The bounded-index requirement in condition (iii) holds
automatically for logistic fitted means and for Gaussian
fitted means with bounded \(\bs Q\).
Condition (iv) controls the collinearity in $\bs G_K$.
The \(K^{-1}\) scale is standard for the
B-spline Gram matrix when the density of index $S$ is bounded
above and away from zero on \([s_-,s_+]\).
Condition (v) restricts the growth of the basis dimension
relative to the labelled sample size.

We additionally impose convergence of the predicted score
\(\bs h_{\mathrm{san},K}
=\bs B_{\bs\theta^\ast,K}\bs\beta_{\mathrm{san},K}\).

\begin{assumption}
\label{ass:population-score-limit}
As \(n\to\infty\), the population predicted scores satisfy
\[
    \E\|\mathcal C\bs h_{\mathrm{san},K}
          -\bs h_{\mathrm{san},\infty}\|^2
    \longrightarrow 0,
\]
where \(\bs h_{\mathrm{san},\infty}\) is a random vector with zero mean and
finite variance, and
\(\bs\Sigma(\bs h_{\mathrm{san},\infty})\)
is positive definite.
\end{assumption}

\begin{theorem}
\label{thm:san-asymptotic}
Suppose that
\Cref{ass:sampling,ass:score-learning,ass:population-score-limit}
hold.
If \(n,N\to\infty\) and
\(\rho_n=n/(n+N)\to\rho\in[0,1)\), then

(i) $\sqrt n
    (\widehat{\bs\theta}_{\mathrm{san},K}-\bs\theta^\ast)
    \rightsquigarrow
    \mathcal N\{\bs0,\bs\Sigma(\bs h_{\mathrm{san},\infty})\}$ and

(ii) $\tr\{\bs\Sigma(\bs h_{\mathrm{san},\infty})\}
    \leq \tr(\bs\Sigma_{\rm lab})$.
\end{theorem}
\Cref{thm:san-asymptotic}(i) shows that the proposed
estimator is asymptotically normal and its asymptotic covariance depends on \(\bs h_{\mathrm{san},\infty}\). The
optimality is achieved when
\(\bs h_{\mathrm{san},\infty}= \bs h^{\mathrm{opt}}\). \Cref{thm:san-asymptotic}(ii) shows that the proposed
estimator has an asymptotic covariance trace no greater
than that of the labelled-only estimator.
This efficiency guarantee holds even when \(S\)
provides a poor approximation to the conditional mean
\(\E(Y\mid f,\bs X)\). The magnitude of the efficiency gain, however, depends
on how well \(\bs h_{\mathrm{san},\infty}\) approximates
\(\bs h^{\mathrm{opt}}\).

 \Cref{cor:variance-consistency} establishes
consistency of the covariance estimator in
\eqref{eq:variance-estimator}.

\begin{corollary}
\label{cor:variance-consistency}
Under the conditions of \Cref{thm:san-asymptotic},
$
    \widehat{\bs\Sigma}_{\mathrm{san},K}
    \xrightarrow{\mathbb P}
    \bs\Sigma(\bs h_{\mathrm{san},\infty}).
$
Consequently, for any fixed nonzero
\(\bs c\in\mathbb R^p\) and \(0<\alpha<1\),
the corresponding Wald interval for
\(\bs c^\top\bs\theta^\ast\) has limiting coverage probability
\(1-\alpha\).
\end{corollary}

\subsection{Case of Sufficient Dimension Reduction}

Although our efficiency guarantee does not require that $\E(Y\mid f,X)$ be represented as a function of the scalar
index $S$, the following proposition
establishes this property under a specific model.

\begin{proposition}
\label{thm:conditional-mean-preservation}
Suppose that, for some measurable function
\(g:\mathbb R\to\mathbb R\) and coefficient vector
\(\bs\alpha\in\mathbb R^p\),
\begin{equation}
    \E\{Y\mid \bs Q\}=g(T),
    \qquad
    T=\bs\alpha^\top\bs Q.
    \label{eq:conditional-single-index}
\end{equation}
Suppose that \(\bs Q\) has an elliptical distribution
with finite second moments and a nonsingular covariance matrix.
Assume that \(A\) is twice continuously differentiable with
\(A''>0\) and that $\E\!\left[
        \|\widetilde{\bs Q}\|
        \{|Y|+|S|\}
    \right]<\infty.$
Finally, suppose that either \(g(T)\) is almost surely constant or
\begin{equation}
    \Cov\{T,g(T)\}\ne0.
    \label{eq:index-nondegeneracy}
\end{equation}
Then
\[
    \E(Y\mid S)
    =\E(Y\mid f,\bs X)
    \quad\text{almost surely}.
\]
\end{proposition}

The single-index assumption
\eqref{eq:conditional-single-index}, ellipticity assumption,
and the nondegeneracy assumption
\eqref{eq:index-nondegeneracy} are standard conditions
for direction recovery in single-index models
\citep{li1989regression}.
Together, they ensure that the augmented
GLM \eqref{eq:augmented-score} recovers the direction of the true index:
the nonintercept component of \(\bs\gamma^\ast\)
equals \(c\bs\alpha\) for some \(c\ne0\).

\begin{remark}
\label{rem:gaussian-single-index}
We illustrate the single-index assumption
\eqref{eq:conditional-single-index} and the ellipticity
of \(\bs Q\) through a nonlinear extension of the stylized
models for modality mismatch and distributional shift
in \cite{ji2025predictions}.
Let the covariates \(\bs X\sim N(0,\Sigma_X)\) and
the auxiliary predictors \(\bs W\sim N(0,\Sigma_W)\) be independent Gaussian vectors.
Consider that the outcome follows a GLM with conditional mean
\[
    \E(Y\mid\bs X,\bs W)
    =
    A'\!\left(
        \bs\delta_X^\top\bs X
        +\bs\delta_W^\top\bs W
    \right).
\]
We further consider that the prediction \(f\) is obtained
from a GLM with  \(\bs X\) and \(\bs W\),
so that
\[
    \tau(f)
    =(A')^{-1}(f)
    =
    \widetilde{\bs\delta}_X^\top\bs X
    +\widetilde{\bs\delta}_W^\top\bs W.
\]
The prediction coefficients
\((\widetilde{\bs\delta}_X,\widetilde{\bs\delta}_W)\)
may differ from
\((\bs\delta_X,\bs\delta_W)\) due to modality mismatch
or distributional shift.
Then \(\bs Q=(\tau(f),\bs X^\top)^\top\) is multivariate
normal and hence elliptically distributed.
Moreover, we show in the Supplementary Material that
\(\E(Y\mid f,\bs X)=g(\bs\alpha^\top\bs Q)\)
for some coefficient vector \(\bs\alpha\) and
univariate function \(g\), justifying assumption \eqref{eq:conditional-single-index}.
\end{remark}

\section{Numerical studies}
\label{sec:simulation}

We compared estimation efficiency and confidence interval coverage across different methods. We considered linear regression for continuous outcomes and logistic regression for binary outcomes, each with an intercept and covariates $\bs X=(X_1,\ldots,X_5)^\top$. We also included $\bs W=(W_1,\ldots,W_{10})^\top$ as a vector of auxiliary predictors. The vectors $\bs X$ and $\bs W$ were generated independently using latent Gaussian copulas, with latent correlations $0.3^{|j-k|}$ between components $j$ and $k$ within each vector, and Student $t$ marginal distributions with $10$ degrees of freedom for $X_1,X_2,X_3,W_1,\ldots,W_5$ and Bernoulli$(0.25)$ marginals for $X_4,X_5,W_6,\ldots,W_{10}$. Each variable was then centred and scaled using its population mean and standard deviation.
Let $T=L/\Var(L)^{1/2}$, where $L=\sum_{j=1}^5X_j+2(W_1+W_2+W_6+W_7)$. We considered two mean specifications for each outcome type, one linear and the other quadratic in $T$ on the link scale. Specifically, let
\[
\begin{aligned}
m_{{\rm G},1}(t)&=3+2t,
&
m_{{\rm G},2}(t)&=3+2(t+t^2-1),
\\
m_{{\rm B},1}(t)&=\operatorname{expit}(3t),
&
m_{{\rm B},2}(t)&=\operatorname{expit}\{6.5+10(t+t^2-1)\},
\end{aligned}
\]
where $\operatorname{expit}(u)=(1+e^{-u})^{-1}$. For $\ell\in\{1,2\}$, we generated the continuous outcomes by $Y=m_{{\rm G},\ell}(T)+\epsilon$, where $\epsilon\sim\mathcal N(0,1)$ was independent of $(\bs X,\bs W)$, and the binary outcomes by
$Y\mid(\bs X,\bs W)\sim\operatorname{Bernoulli}\{m_{{\rm B},\ell}(T)\}$.
For each scenario, we considered $n=600$ labelled units and $N\in\{500,1000,2000,4000\}$ unlabelled units. We used $1000$ Monte Carlo replications for each setting.

To generate the predictions $f$, we trained a random forest on an external sample of size $5000$, independent of the analysis sample. All random forests were fitted using \texttt{scikit-learn} with $100$ trees and default hyperparameters. For binary outcomes, $f$ was the predicted probability of $Y=1$. To compute the logit transformation for SAN-PPI, probabilities were clipped to $[10^{-6},1-10^{-6}]$. Once trained, the prediction model was held fixed across Monte Carlo replications.
We considered two training scenarios. Under modality mismatch, the training sample came from the study population, but the prediction model used only $\bs W$ as input. Under distributional shift, the prediction model used both $\bs X$ and $\bs W$, but the conditional distribution of $Y$ given $(\bs X,\bs W)$ differed between the training and study populations. Details of the distributional shift setting are provided in the Supplementary Material.

The proposed estimator used a cubic B-spline basis with $K=4$.
We compared it with the labelled-only estimator
and the methods of
\citet{angelopoulos2023prediction,angelopoulos2023ppi++}, \citet{gan2024prediction}, \citet{miao2025assumption}, \citet{song2023general} and \citet{ji2025predictions}.
We also included an estimator that used five-fold cross-fitting
and random forests to directly estimate $\E(Y\mid f,\bs X)$ and construct
the predicted score in \eqref{eq:h-full} without dimension
reduction.
We evaluated the root mean squared error (RMSE) of the full coefficient vector and the empirical coverage of $95\%$ Wald intervals for each coefficient.

\begin{figure}[t]
\centering
\includegraphics[width=\textwidth]{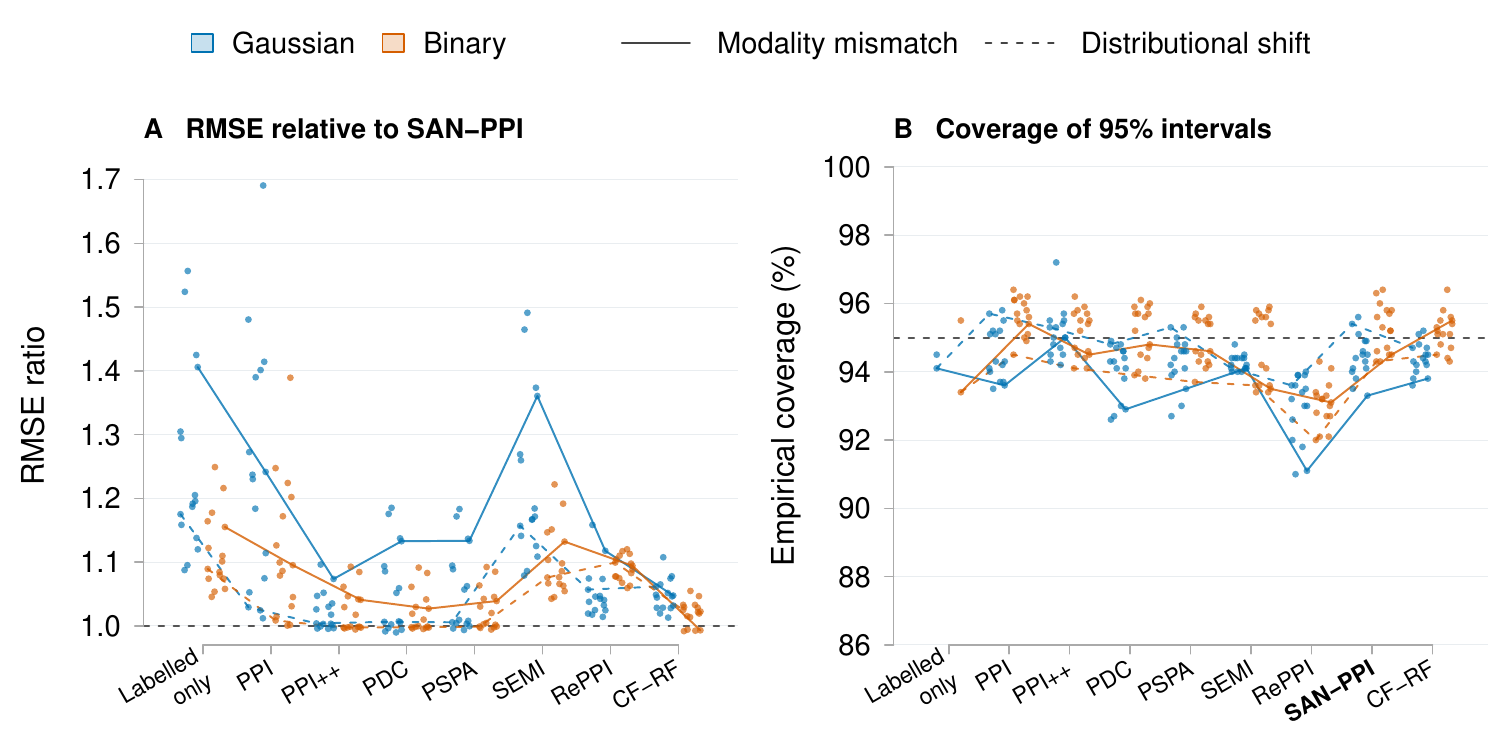}

\caption[RMSE ratios and empirical coverage]{
RMSE ratios and empirical coverage for $n=600$ and $N\in\{500,1000,2000,4000\}$ under two mean specifications and two outcome types, based on $1000$ Monte Carlo replications per setting.
(A) RMSE of the full coefficient vector relative to SAN-PPI, $\mathrm{RMSE}/\mathrm{RMSE}_{\mathrm{SAN}}$.
(B) Empirical coverage of $95\%$ Wald intervals for the coefficient of $X_1$.
Points represent individual settings, while lines connect results for $N=2000$ under the second mean specification.
Blue and orange denote Gaussian and binary outcomes, respectively; solid and dashed lines denote modality mismatch and distributional shift, respectively.
Grey dashed horizontal lines indicate one in panel A and $95\%$ in panel B.
Methods: PPI \citep{angelopoulos2023prediction}, PPI++ \citep{angelopoulos2023ppi++}, PDC \citep{gan2024prediction}, PSPA \citep{miao2025assumption}, SEMI \citep{song2023general}, RePPI \citep{ji2025predictions}, SAN-PPI (the proposed estimator), CF--RF (the cross-fitted random-forest estimator).
}
\label{fig:simulation-summary}
\end{figure}

Figure~\ref{fig:simulation-summary} compares the RMSE of the competing estimators relative to the proposed estimator and the empirical coverage of $95\%$ Wald intervals for the coefficient of $X_1$ across all settings and outcome types. The minimum RMSE ratio relative to the proposed estimator was approximately $0.990$ (panel A), indicating that the proposed estimator was comparable to the best-performing estimator among those considered in terms of RMSE, while its confidence intervals achieved coverage close to the nominal $95\%$ level (panel B). The confidence intervals of \citet{ji2025predictions} show some undercoverage, possibly due to the relatively small labelled sample size. We found results for $n=300$ were similar to those for $n=600$.

The Supplementary Material provides comparisons with an approximate
oracle estimator, which uses a large independent simulated sample
to estimate \(\E(Y\mid f,\bs X)\) in the optimal predicted score
\(\bs h^{\mathrm{opt}}\).
It also reports sensitivity analyses for the B-spline basis dimension
\(K\in\{5,6,7\}\) and coverage results for the remaining coefficients. The proposed estimator has negligible empirical bias and RMSE comparable to that of the approximate oracle estimator, performs similarly across choices of $K$, and shows similar coverage patterns for the remaining coefficients. Marginal distribution plots of $\tau(f)$, also provided in the Supplementary Material, show pronounced skewness in some settings, particularly under the second mean specification. The efficiency gains in these settings suggest that the proposed method can remain effective beyond the ellipticity assumption used to justify dimension reduction.

\section*{Declaration of the use of generative AI and AI-assisted technologies}

During the preparation of this manuscript, the authors used OpenAI Codex to assist with coding and mathematical proofs. The authors subsequently reviewed and edited the content as necessary and assume full responsibility for the content of the publication.

\section*{Supplementary material}

The Supplementary Material follows the references and includes proofs, technical lemmas, and additional simulation results.

\FloatBarrier

\bibliographystyle{abbrvnat}
\bibliography{refs}

\clearpage
\section*{Supplementary material}
\addcontentsline{toc}{section}{Supplementary material}
% Reset supplementary counters; unique hyperlink anchors prevent collisions
% with the main article while preserving the original S-prefixed numbering.
\setcounter{section}{0}
\renewcommand{\thesection}{S\arabic{section}}
\renewcommand{\theHsection}{supp.\arabic{section}}
\setcounter{equation}{0}
\renewcommand{\theequation}{S\arabic{equation}}
\renewcommand{\theHequation}{supp.\arabic{equation}}
\setcounter{figure}{0}
\renewcommand{\thefigure}{S\arabic{figure}}
\renewcommand{\theHfigure}{supp.\arabic{figure}}
\setcounter{table}{0}
\renewcommand{\thetable}{S\arabic{table}}
\renewcommand{\theHtable}{supp.\arabic{table}}
\setcounter{theorem}{0}
\renewcommand{\thetheorem}{S\arabic{theorem}}
\renewcommand{\theHtheorem}{supp.\arabic{theorem}}
\setcounter{lemma}{0}
\renewcommand{\thelemma}{S\arabic{lemma}}
\renewcommand{\theHlemma}{supp.\arabic{lemma}}
\setcounter{corollary}{0}
\renewcommand{\thecorollary}{S\arabic{corollary}}
\renewcommand{\theHcorollary}{supp.\arabic{corollary}}
\setcounter{proposition}{0}
\renewcommand{\theproposition}{S\arabic{proposition}}
\renewcommand{\theHproposition}{supp.\arabic{proposition}}
\setcounter{assumption}{0}
\renewcommand{\theassumption}{S\arabic{assumption}}
\renewcommand{\theHassumption}{supp.\arabic{assumption}}
\setcounter{remark}{0}
\renewcommand{\theremark}{S\arabic{remark}}
\renewcommand{\theHremark}{supp.\arabic{remark}}

Section~\ref{sec:simulation-supplement} provides details of the distributional shift setting
and additional simulation results referred to in Section~\ref{sec:simulation} of the main paper,
including comparisons with an approximate oracle estimator, bias and interval coverage,
sensitivity to the spline basis dimension, and marginal distributions of the transformed prediction.
Section~\ref{supp:theory-proofs} contains proofs of the theoretical results in the main paper
and supporting technical lemmas.
It also specifies the B-spline basis used in Assumption~\ref{ass:score-learning}(iii)
and provides the derivation of the conditional mean representation in
Remark~\ref{rem:gaussian-single-index} of the main paper.

\section{Additional simulation results}
\label{sec:simulation-supplement}
\subsection{Additional simulation details}
\label{sec:details}

Under distributional shift, $(\bs X,\bs W)$ had the same distribution
in the training and study populations. In the training population,
outcomes were generated using
\begin{equation}
 T_{\rm src}=\frac{L_{\rm src}}{\Var(L_{\rm src})^{1/2}},\qquad
 L_{\rm src}=\sum_{j=1}^5X_j+2W_1+2W_2+W_6-W_7
 \label{eq:source-index}
\end{equation}
in place of $T$ in $m_{{\rm G},\ell}(T)$ and $m_{{\rm B},\ell}(T)$, for $\ell=1,2$.

For the proposed estimator, the knots of the cubic B-spline basis were selected in each Monte Carlo replication from the estimated indices $\{\widehat S_i\}_{i=1}^{n+N}$. The boundary knots were set to $\min_{1\le i\le n+N}\widehat S_i$ and $\max_{1\le i\le n+N}\widehat S_i$, and the $K-4$ interior knots were placed at the empirical $j/(K-3)$ quantiles of these indices, $j=1,\ldots,K-4$.

For the cross-fitted random-forest estimator, we partitioned the labelled sample into five folds $\mathcal I_1,\ldots,\mathcal I_5$. Let $\widehat m^{(-k)}$ denote the random forest fitted using the labelled observations outside $\mathcal I_k$, with $(f,\bs X)$ as predictors. We constructed

$$
\widehat m_i=
\begin{cases}
\widehat m^{(-k)}(f_i,\bs X_i),
& i\in\mathcal I_k,\quad k=1,\ldots,5,\\[2mm]
\displaystyle\frac15\sum_{k=1}^5\widehat m^{(-k)}(f_i,\bs X_i),
& i=n+1,\ldots,n+N.
\end{cases}
$$
The predicted score was
$$
\widehat{\bs h}_i
=(1-\rho_n)\widetilde{\bs X}_i
(\widehat m_i-\widehat\mu_i),
$$
which was substituted into the one-step estimator in \eqref{eq:one-step-ppi}.

For the method of \citet{ji2025predictions}, we used linear regression for Gaussian outcomes and $L_2$-penalized logistic regression for binary outcomes to construct the imputed score. Both regression models used the basis $(1,f,\bs X^\top)^\top$. The same basis was used for the method of \citet{song2023general}.

\subsection{Comparison with the approximate oracle estimator}
\label{sec:performance}
To construct the approximate oracle estimator, we fitted a random forest to estimate $\E(Y\mid f,\bs X)$ using $12000$ additional labelled observations drawn independently from the study population. The random forest used the same hyperparameters as those described in the main paper. The resulting estimate, denoted by $\widehat m_{\rm ref}(f,\bs X)$, was held fixed across Monte Carlo replications.
We constructed
\begin{equation}
 \widehat{\bs h}_{\rm ref}(f,\bs X)
 =(1-\rho_n)\widetilde{\bs X}
 \{\widehat m_{\rm ref}(f,\bs X)-A'(\widetilde{\bs X}^{\top}\thetalab)\},
 \label{eq:reference-score}
\end{equation}
and substituted this score into the one-step estimator in
\eqref{eq:one-step-ppi}.
This benchmark was approximate because the
conditional mean was estimated from a finite additional sample.

\begin{figure}[!htbp]
\centering\includegraphics[width=\textwidth]{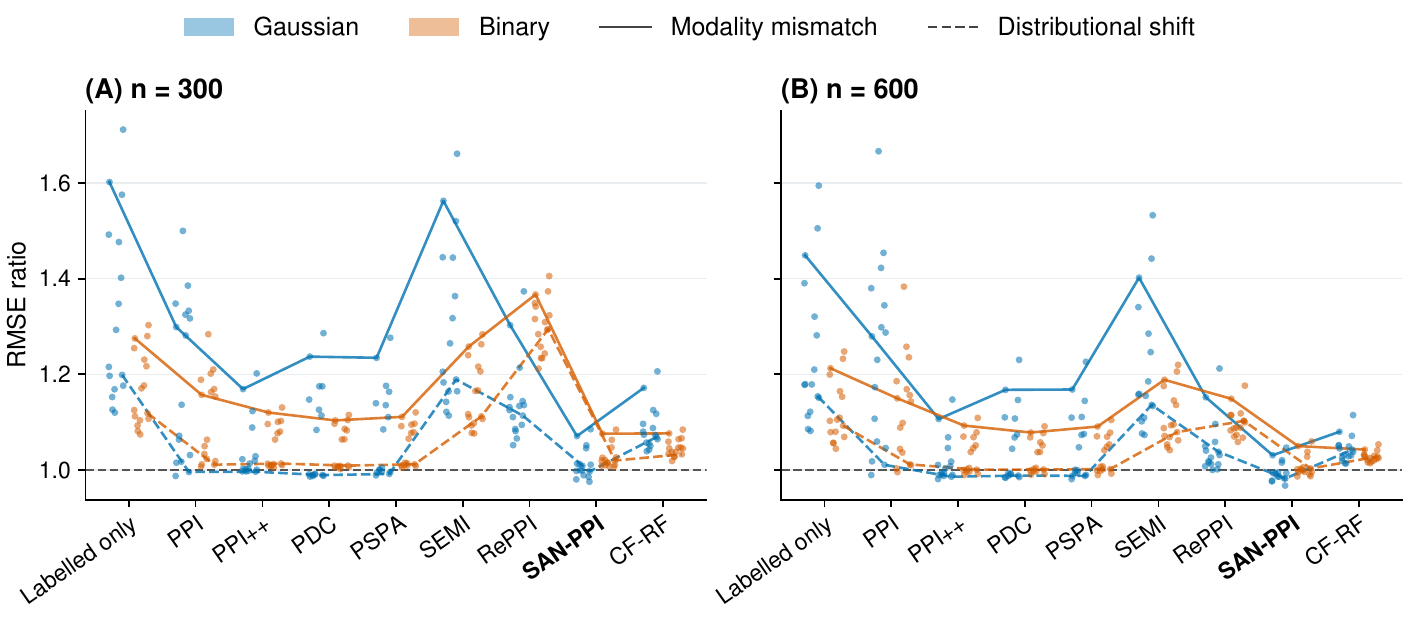}

\caption[Comparison with the approximate oracle estimator]{
RMSE ratios of the full coefficient vector relative to the approximate oracle estimator, $\mathrm{RMSE}/\mathrm{RMSE}_{\rm ref}$, for $N\in\{500,1000,2000,4000\}$ under two mean specifications and two outcome types, based on $1000$ Monte Carlo replications per setting.
(A) $n=300$. (B) $n=600$.
Points represent individual settings, while lines connect results for $N=2000$ under the second mean specification.
Blue and orange denote Gaussian and binary outcomes, respectively; solid and dashed lines denote modality mismatch and distributional shift, respectively.
Grey dashed horizontal lines indicate one.
Method labels are as in Fig.~\ref{fig:simulation-summary} of the main paper.
}
\label{fig:reference}
\end{figure}

Figure~\ref{fig:reference} shows the RMSE of the competing estimators relative to that of the approximate oracle estimator. Overall, the RMSE of the proposed estimator was closest to that of the approximate oracle estimator.

\subsection{Bias and interval coverage}

\begin{figure}[!tbp]
\centering\includegraphics[width=\textwidth]{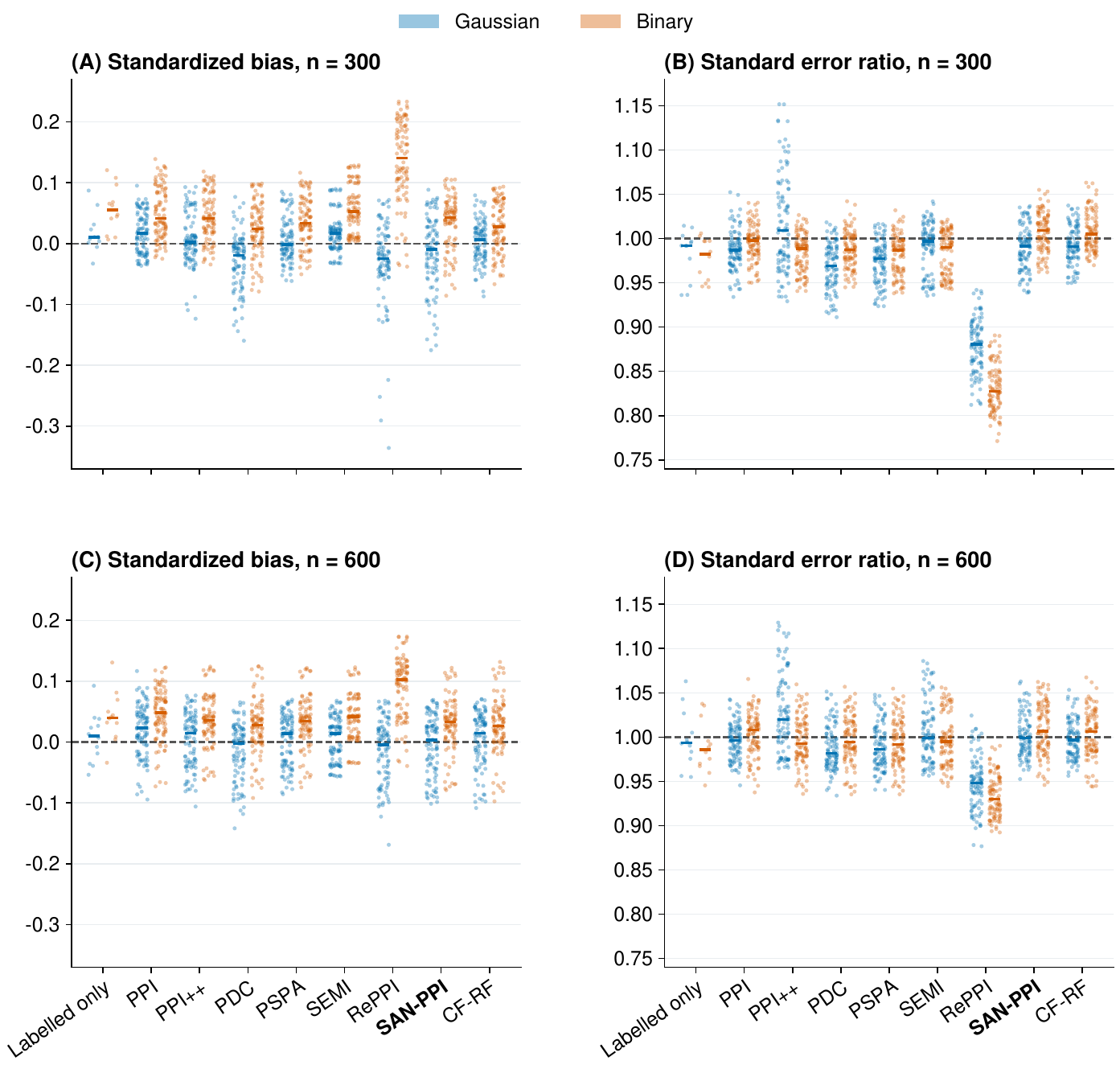}

\caption[Standardized bias and standard-error ratios]{
Standardized bias and standard-error ratios for $n=300$ (top) and $n=600$ (bottom), based on $1000$ Monte Carlo replications per setting.
(A, C) Standardized bias, $(\overline{\widehat\theta}_j-\theta_j^\ast)/s_j$.
(B, D) Standard-error ratio, $\overline{\widehat{\rm se}}_j/s_j$.
Here $\overline{\widehat\theta}_j$ and $\overline{\widehat{\rm se}}_j$ are the Monte Carlo averages of the coefficient estimates and their estimated standard errors, respectively, and $s_j$ is the empirical standard deviation of $\widehat\theta_j$.
Points represent individual coefficient-setting combinations; short horizontal segments indicate medians.
Blue and orange denote Gaussian and binary outcomes, respectively.
Grey dashed horizontal lines indicate zero in panels A and C and one in panels B and D.
Method labels are as in Fig.~\ref{fig:simulation-summary} of the main paper.
}

\label{fig:bias}
\end{figure}

\begin{figure}[!tbp]
\centering\includegraphics[width=.90\textwidth]{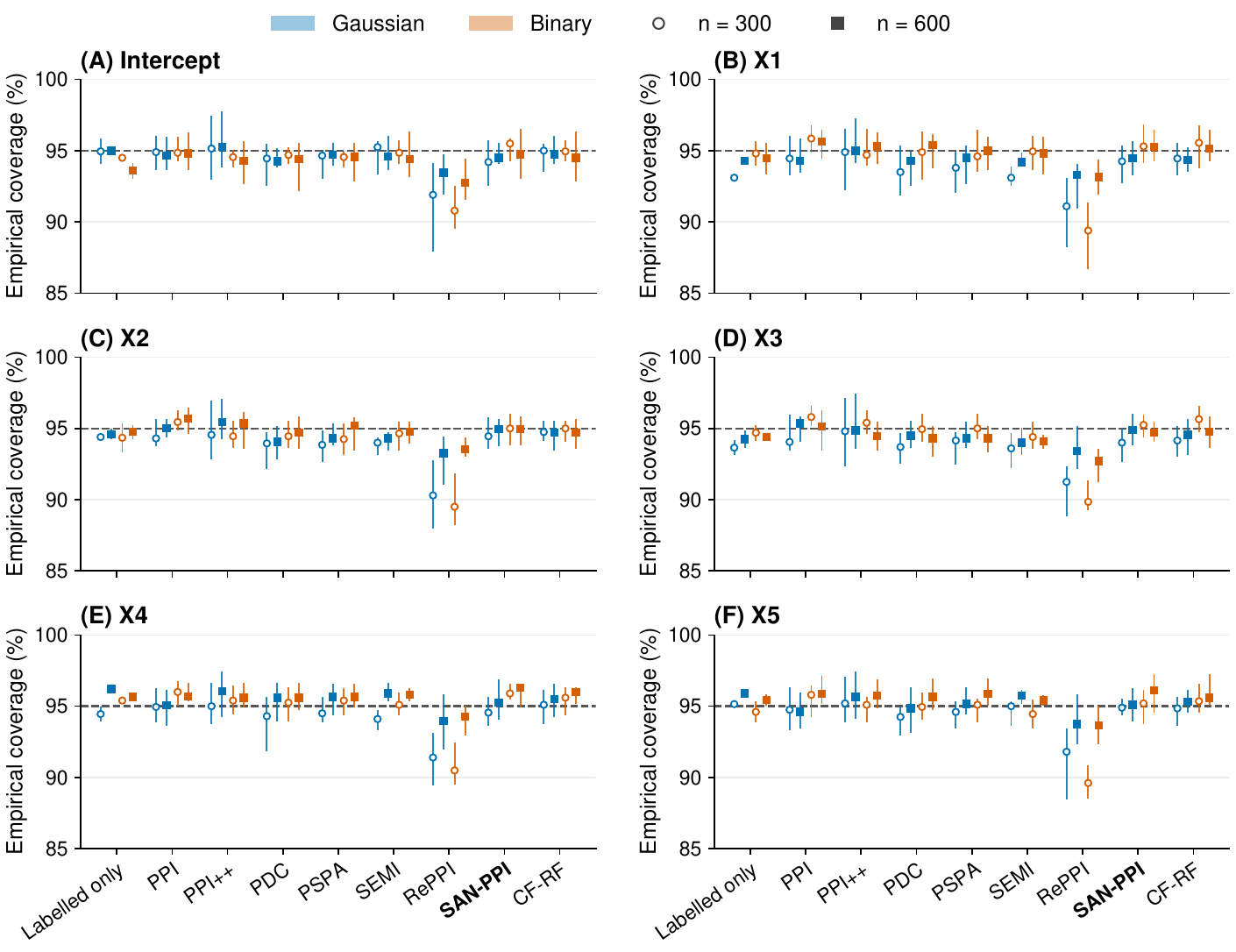}

\caption[Empirical coverage for the remaining coefficients]{
Empirical coverage of $95\%$ Wald intervals for the intercept and five covariate coefficients at $n=300$ and $n=600$, based on $1000$ Monte Carlo replications per setting.
Symbols indicate median coverage across $16$ settings; vertical bars indicate the minimum-to-maximum ranges.
Blue and orange denote Gaussian and binary outcomes, respectively; open circles and filled squares denote $n=300$ and $n=600$, respectively.
Grey dashed horizontal lines indicate $95\%$.
Method labels are as in Fig.~\ref{fig:simulation-summary} of the main paper.
}

\label{fig:coefficient-coverage}
\end{figure}

Figure~\ref{fig:bias} shows the standardized bias and standard error ratios of the competing estimators, while Fig.~\ref{fig:coefficient-coverage} shows the empirical coverage for each coefficient. Overall, the proposed estimator had small empirical bias and estimated standard errors close to the empirical standard deviations, with coverage near the nominal level.

\subsection{Sensitivity to the dimension of the spline basis}
\label{sec:k}
\begin{figure}[!tbp]
\centering\includegraphics[width=.92\textwidth]{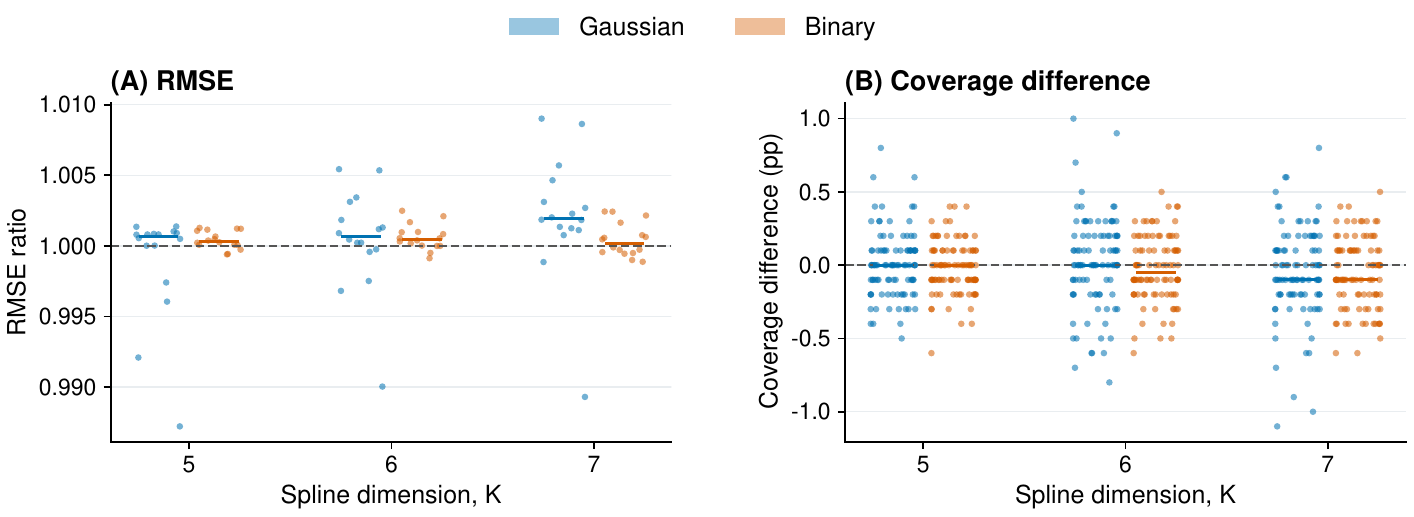}

\caption[Sensitivity to the spline basis dimension]{
RMSE ratios and coverage differences for spline dimensions $K\in\{5,6,7\}$ relative to $K=4$, at $n=600$.
(A) RMSE ratio for the full coefficient vector, $\mathrm{RMSE}_K/\mathrm{RMSE}_4$.
(B) Coverage difference, $100(\mathrm{CP}_{j,K}-\mathrm{CP}_{j,4})$, in percentage points, where $\mathrm{CP}_{j,K}$ is the empirical coverage of $95\%$ Wald intervals for coefficient $j$ with spline dimension $K$.
Points represent individual settings in panel A and coefficient-setting combinations in panel B; short horizontal segments indicate medians.
Blue and orange denote Gaussian and binary outcomes, respectively.
Grey dashed horizontal lines indicate one in panel A and zero in panel B.
}

\label{fig:k}
\end{figure}

Figure~\ref{fig:k} shows the RMSE ratios and coefficient-specific coverage differences relative to $K=4$ for $K\in\{5,6,7\}$. The RMSE ratios remained close to one, ranging from $0.987$ to $1.009$, while the absolute coverage differences were less than $1.1$ percentage points. These results show that the performance of the proposed estimator was insensitive to the choice of $K$ over the values considered.

\subsection{Marginal distributions of the transformed prediction}
\label{sec:tau}
\begin{figure}[!tbp]
\centering\includegraphics[width=.96\textwidth]{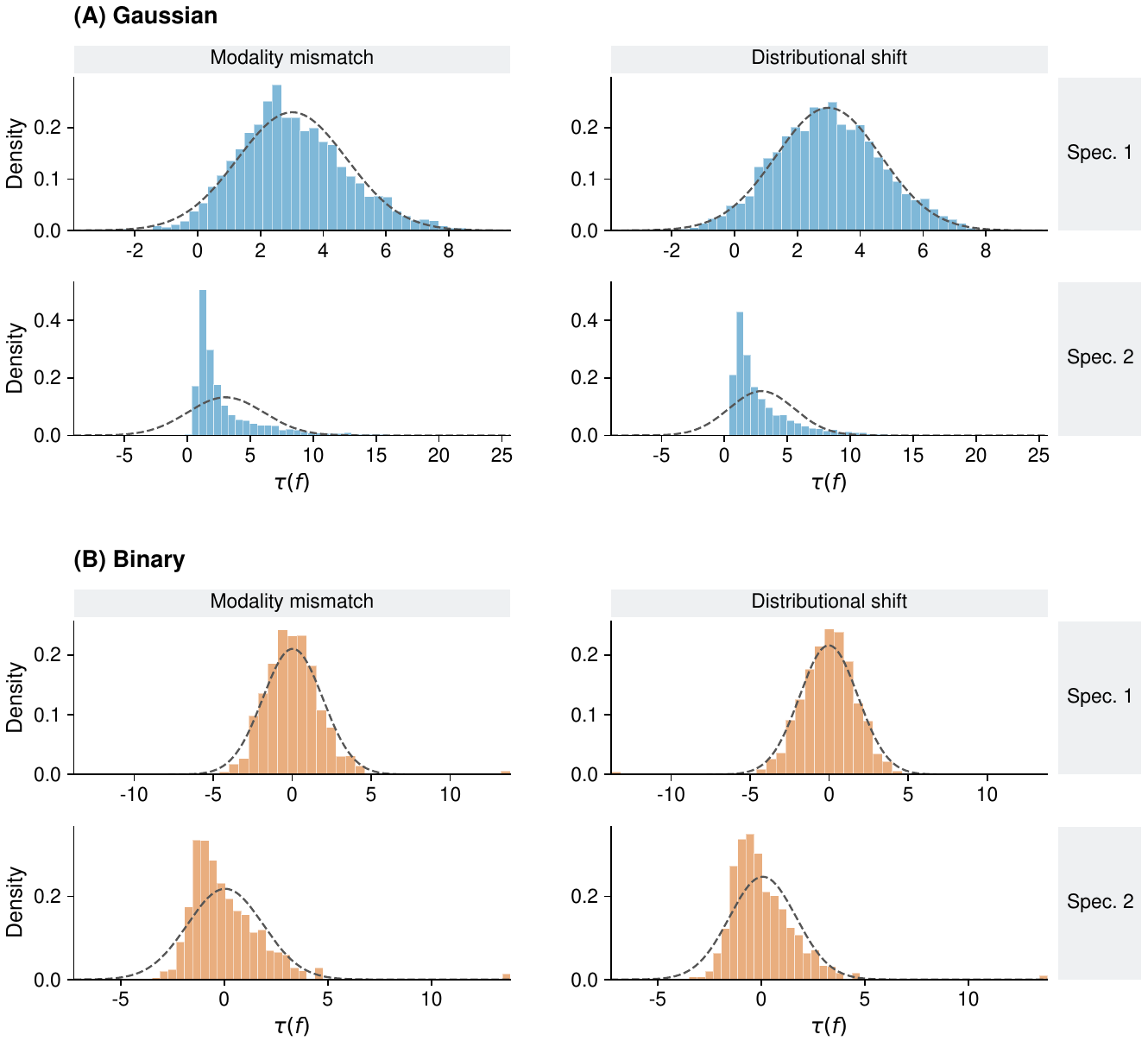}

\caption[Marginal distributions of the transformed prediction]{
Marginal distributions of the transformed prediction $\tau(f)$ for Gaussian (A) and binary (B) outcomes.
Columns correspond to modality mismatch (left) and distributional shift (right); rows correspond to the first (Spec.~1) and second (Spec.~2) mean specifications.
Grey dashed curves indicate normal densities matched to the empirical means and variances.
}
\label{fig:tau}
\end{figure}

Figure~\ref{fig:tau} shows pronounced right skewness in the transformed predictions under the second mean specification.

\clearpage
\section{Proofs and supplementary derivations}
\label{supp:theory-proofs}

We use the notation and assumptions of the main paper. Write
\(\|A\|_{m,2}=(\mathbb E_m\|A\|_F^2)^{1/2}\), \(m\in\{n,N\}\).
The norm \(\|\cdot\|_{\rm op}\) is the matrix operator norm.
\allowdisplaybreaks[1]

\subsection{Proof of Proposition~\ref{prop:beta-closed-form}}

\begin{proof}
For a fixed \(\bs\beta\), let
\[
    \bs v=(\mathcal C\bs B_{\bs\theta^\ast,K})\bs\beta,
    \qquad
    \bs u=\mathcal C\big\{\bs\psi_{\bs\theta^\ast}(Y,\bs X)-\bs B_{\bs\theta^\ast,K}\bs\beta\big\}.
\]
By \eqref{eq:ppi-h} and the cyclic property of the trace,
\begin{align*}
    \tr\{\bs\Sigma(\bs B_{\bs\theta^\ast,K}\bs\beta)\}
    &=
    \E\tr\bigl(\bs J^{-1}\bs u\bs u^\top\bs J^{-\top}\bigr)
    +\frac{\rho}{1-\rho}
     \E\tr\bigl(\bs J^{-1}\bs v\bs v^\top\bs J^{-\top}\bigr) \\
    &=
    \E(\bs u^\top\bs M\bs u)
    +\frac{\rho}{1-\rho}\E(\bs v^\top\bs M\bs v) \\
    &=
    \tr(\bs\Sigma_{\rm lab})
    -2\E\{\bs v^\top\bs M(\mathcal C\bs\psi_{\bs\theta^\ast})\}
    +\frac{1}{1-\rho}\E(\bs v^\top\bs M\bs v) \\
    &=
    \tr(\bs\Sigma_{\rm lab})
    -2\bs\beta^\top\bs d_K
    +\frac{\bs\beta^\top\bs G_K\bs\beta}{1-\rho}.
\end{align*}

By \eqref{eq:G-d}, \(\bs G_K\) is positive semidefinite
and hence positive definite under the nonsingularity assumption.
The trace is strictly convex in \(\bs\beta\) and therefore has a unique minimizer $\bs\beta_{\mathrm{san},K}=(1-\rho)\bs G_K^{-1}\bs d_K$.
Substituting this minimizer into the trace criterion gives
\begin{align*}
 &\tr(\bs\Sigma_{\rm lab})
   -\tr\{\bs\Sigma(\bs B_{\bs\theta^\ast,K}\bs\beta_{\mathrm{san},K})\}\\
 &\quad=2\bs\beta_{\mathrm{san},K}^\top\bs d_K
       -\frac{\bs\beta_{\mathrm{san},K}^\top\bs G_K
                       \bs\beta_{\mathrm{san},K}}{1-\rho}\\
 &\quad=(1-\rho)\bs d_K^\top\bs G_K^{-1}\bs d_K\geq0.
\end{align*}

\end{proof}

\subsection{B-spline basis}
\label{supp:bspline}

Fix a degree \(r\geq2\) and a dimension \(K\geq r+1\).
Let \(s_-=\xi_0<\xi_1<\cdots<\xi_{K-r-1}<\xi_{K-r}=s_+\)
(no interior knots when \(K=r+1\)), and consider the knot sequence
\[
 (\tau_1,\ldots,\tau_{K+r+1})
 =\bigl(\underbrace{s_-,\ldots,s_-}_{r+1},
        \xi_1,\ldots,\xi_{K-r-1},
        \underbrace{s_+,\ldots,s_+}_{r+1}\bigr).
\]
Write \(\ell_j=\xi_j-\xi_{j-1}\), \(j=1,\ldots,K-r\), and
\(\ell_{\min}=\min_j\ell_j\), \(\ell_{\max}=\max_j\ell_j\).
The knots are quasi-uniform: \(\ell_{\max}/\ell_{\min}\leq\kappa\)
for a constant \(\kappa\) independent of \(K\).
For \(s\in[s_-,s_+]\), let
\[
 \bs b_K(s)=\{b_{1,K}(s),\ldots,b_{K,K}(s)\}^\top
\]
denote the degree-\(r\) B-spline basis on this knot sequence
\citep{deboor1993basics}.

\begin{lemma}
\label{supp:lem:spline-bounds}
There is a constant \(C\) depending only on \(r\), \(\kappa\) and
\(s_+-s_-\) such that, for all \(s,t\in[s_-,s_+]\),
\begin{equation}
 \|\bs b_K(s)\|\leq1,\qquad
 \|\bs b'_K(s)\|\leq CK,\qquad
 \|\bs b'_K(t)-\bs b'_K(s)\|\leq CK^2|t-s|,
 \label{supp:eq:spline-bounds}
\end{equation}
with one-sided derivatives at \(s_-\) and \(s_+\).
\end{lemma}

\begin{proof}
Since $b_{k,K}(s)\geq0$ and $\sum_{k=1}^Kb_{k,K}(s)=1$, we have
\[
    \|\bs b_K(s)\|^2
    =\sum_{k=1}^K b_{k,K}(s)^2
    \leq\sum_{k=1}^K b_{k,K}(s)=1.
\]

Let \(N_{k,q}\) denote the degree-\(q\) B-spline with index \(k\) on the given knot
sequence, so that \(b_{k,K}=N_{k,r}\). For \(q\geq1\), its derivative is
\[
    N_{k,q}'(s)
    =\frac{q N_{k,q-1}(s)}{\tau_{k+q}-\tau_k}
     -\frac{q N_{k+1,q-1}(s)}{\tau_{k+q+1}-\tau_{k+1}},
\]
where \(N_{k,q-1}\) is supported on \([\tau_k,\tau_{k+q}]\) and \(0/0\) is taken to be zero. Since \(0\leq N_{k,q}\leq1\), applying the
recurrence once and twice gives, on each open knot interval,
\[
    |N_{k,r}'(s)|\leq\frac{2r}{\ell_{\min}},
    \qquad
    |N_{k,r}''(s)|\leq\frac{4r(r-1)}{\ell_{\min}^2}.
\]
Quasi-uniformity implies \(\ell_{\min}\geq c/K\).
By the local support property of B-splines, at most \(r+1\)
basis functions are nonzero at any given point \(s\).
Combining these properties with the preceding componentwise
derivative bounds gives
\(\|\bs b_K'(s)\|\leq CK\) and
\(\|\bs b_K''(s)\|\leq CK^2\).
Hence, for any \(s,t\in[s_-,s_+]\),
\[
    \|\bs b_K'(t)-\bs b_K'(s)\|
    =\left\|\int_s^t \bs b_K''(u)\,du\right\|
    \leq CK^2|t-s|.
\]
\end{proof}

\subsection{Proof of Theorem~\ref{thm:san-asymptotic}}

\label{supp:theorem-proof}

Here and below, \(\bs I_d\) denotes the \(d\times d\) identity matrix.
The symbol \(C\) denotes a positive constant independent of \(n,N,K\),
whose value may change from line to line.

Let \(\Delta_n=\mathbb E_N-\mathbb E_n\).
For any fixed square-integrable random matrix \(H\) observed in both
samples, \(\Delta_n H=\Delta_n\mathcal C H\). Independence gives
\begin{equation}
 \begin{aligned}
 \E\|\sqrt n\Delta_n H\|_F^2
 &=n\E\|\mathbb E_N\mathcal C H\|_F^2
   +n\E\|\mathbb E_n\mathcal C H\|_F^2\\
 &\quad-2n\E\langle\mathbb E_N\mathcal C H,
                        \mathbb E_n\mathcal C H\rangle_F\\
 &=(1+n/N)\E\|\mathcal C H\|_F^2,
 \end{aligned}
 \label{supp:eq:two-sample-moment}
\end{equation}
where the cross term is zero because the two centred sample means are
independent and have mean zero.

Define
\begin{equation}
 \bs h^0_K=\frac{1-\rho_n}{1-\rho}\bs h_{\mathrm{san},K}.
 \label{supp:eq:comparator}
\end{equation}

We use the following two technical lemmas. Their proofs are given in
Section~\ref{supp:technical-proofs}.

\begin{lemma}
\label{supp:lem:normal-equations}
Suppose that Assumptions~\ref{ass:sampling} and
\ref{ass:score-learning} hold. Let \(K=K_n\), and suppose that
\(n,N\to\infty\) with \(\rho_n=n/(n+N)\to\rho\in[0,1)\). Then
\begin{equation*}
\begin{split}
 \|\bs G_K^{-1/2}\widehat{\bs G}_K\bs G_K^{-1/2}
       -\bs I_{K+1}\|_{\rm op}
   &=O_{\mathbb P}\{(K^3/n)^{1/2}\},\\
 \|\bs G_K^{1/2}(\widehat{\bs\beta}_{\mathrm{san},K}
                     -\tfrac{1-\rho_n}{1-\rho}
                       \bs\beta_{\mathrm{san},K})\|
   &=O_{\mathbb P}\{(K^3/n)^{1/2}\}.
\end{split}
\end{equation*}
In particular, \(\widehat{\bs G}_K\) is nonsingular with probability
tending to one.
\end{lemma}

\begin{lemma}
\label{supp:lem:score-replacement}
Under the conditions of Lemma~\ref{supp:lem:normal-equations}, let
\(\bs\delta_K=\widehat{\bs h}_{\mathrm{san},K}-\bs h^0_K\).
Then
\begin{align}
 \sqrt n\,\Delta_n\bs\delta_K
     &=o_{\mathbb P}(1),
       \label{supp:eq:mean-replacement}\\
 \tr\{\widehat{\bs V}_n(\bs\delta_K)\}
  +\tr\{\widehat{\bs V}_N(\bs\delta_K)\}
     &=o_{\mathbb P}(1).
       \label{supp:eq:variance-replacement}
\end{align}
\end{lemma}

\begin{proof}[Proof of Theorem~\ref{thm:san-asymptotic}]
\smallskip\noindent\textit{Step 1: }
Write \(\widehat{\bs\psi}=\bs\psi_{\widehat{\bs\theta}_{\rm lab}}\).
We prove
\begin{equation}
 \widehat{\bs\theta}_{\rm lab}-\bs\theta^\ast=O_{\mathbb P}(n^{-1/2}),
 \qquad
 \sqrt n(\widehat{\bs\theta}_{\rm lab}-\bs\theta^\ast)
 =\bs J^{-1}\sqrt n\mathbb E_n\bs\psi^\ast+o_{\mathbb P}(1),
 \label{supp:eq:labelled-expansion}
\end{equation}
and
\begin{equation}
 \|\widehat{\bs J}-\bs J\|_F
 +\|\widehat{\bs\psi}-\bs\psi^\ast\|_{n,2}
 =O_{\mathbb P}(n^{-1/2}).
 \label{supp:eq:target-rates}
\end{equation}
On
\(\mathcal U_\theta\),
Assumption~\ref{ass:score-learning}(ii) gives
\begin{align*}
 \E\|\bs\psi^\ast\|^2
 &\leq2\E\{\|\widetilde{\bs X}\|^2(|Y|^2+D^2)\}\\
 &\leq2\{\E(\|\widetilde{\bs X}\|^4|Y|^4)\}^{1/2}
      +2\{\E(\|\widetilde{\bs X}\|^4D^4)\}^{1/2}<\infty.
\end{align*}
For
\(\dot{\bs\psi}_{\bs\theta} =-\widetilde{\bs X}(\partial_{\bs\theta}\mu_{\bs\theta})^\top\),
the same assumption gives
\begin{align*}
 \E\sup_{\bs\theta\in\mathcal U_\theta}
       \|\dot{\bs\psi}_{\bs\theta}\|_F
 &\leq\E(\|\widetilde{\bs X}\|D)
 \leq\{\E(\|\widetilde{\bs X}\|^4D^4)\}^{1/4}<\infty,\\
 \|\dot{\bs\psi}_{\bs\theta}-\dot{\bs\psi}_{\bs\theta^\ast}\|_F
 &\leq\|\widetilde{\bs X}\|D\|\bs\theta-\bs\theta^\ast\|.
\end{align*}

Let \(\lambda=\lambda_{\min}(\bs J)>0\). Choose \(\delta>0\) such that
\[
    \{\bs\theta\in\mathbb R^p:
      \|\bs\theta-\bs\theta^\ast\|\leq\delta\}
    \subseteq\mathcal U_\theta,
    \qquad
    \delta\,\E(\|\widetilde{\bs X}\|D)\leq\lambda/4.
\]
Using \(\bs J=-\E\dot{\bs\psi}_{\bs\theta^\ast}\) and the preceding
derivative bound, we obtain
\[
\begin{aligned}
    \sup_{\|\bs\theta-\bs\theta^\ast\|\leq\delta}
    \|-\mathbb E_n\dot{\bs\psi}_{\bs\theta}-\bs J\|_{\rm op}
    &\leq
    \|(\mathbb E_n-\E)\dot{\bs\psi}_{\bs\theta^\ast}\|_F
    +\sup_{\|\bs\theta-\bs\theta^\ast\|\leq\delta}
      \mathbb E_n\|\dot{\bs\psi}_{\bs\theta}
                   -\dot{\bs\psi}_{\bs\theta^\ast}\|_F \\
    &\leq
    \|(\mathbb E_n-\E)\dot{\bs\psi}_{\bs\theta^\ast}\|_F
    +\delta\,\mathbb E_n(\|\widetilde{\bs X}\|D).
\end{aligned}
\]

By the law of large numbers, the right-hand side converges in
probability to \(\delta\E(\|\widetilde{\bs X}\|D)\leq\lambda/4\).
Thus, with probability tending to one, the supremum is at most
\(\lambda/2\), and
\[
    \inf_{\|\bs\theta-\bs\theta^\ast\|\leq\delta}
    \lambda_{\min}(-\mathbb E_n\dot{\bs\psi}_{\bs\theta})
    \geq \lambda/2.
\]

We now show that
\(\widehat{\bs\theta}_{\rm lab}-\bs\theta^\ast=o_{\mathbb P}(1)\).
Consider
\[
    L_n(\bs\theta)
    =\mathbb E_n\{A(\widetilde{\bs X}^\top\bs\theta)
      -Y\widetilde{\bs X}^\top\bs\theta\},
    \qquad
    \nabla L_n(\bs\theta)=-\mathbb E_n\bs\psi_{\bs\theta}.
\]
Since \(\E\bs\psi^\ast=\bs0\) and \(\E\|\bs\psi^\ast\|^2<\infty\),
\(\mathbb E_n\bs\psi^\ast=O_{\mathbb P}(n^{-1/2})\).
Thus, with probability tending to one, $\|\mathbb E_n\bs\psi^\ast\|<\lambda\delta/4$. Define
\[
    \mathcal E_n
    =
    \left\{
        \|\mathbb E_n\bs\psi^\ast\|<\frac{\lambda\delta}{4},
        \quad
        \inf_{\|\bs\theta-\bs\theta^\ast\|\leq\delta}
        \lambda_{\min}(-\mathbb E_n\dot{\bs\psi}_{\bs\theta})
        \geq\frac{\lambda}{2}
    \right\}.
\]
We have shown that \(\mathbb P(\mathcal E_n)\to1\).
On \(\mathcal E_n\), for every \(\|\bs u\|=\delta\),
\begin{align*}
    L_n(\bs\theta^\ast+\bs u)-L_n(\bs\theta^\ast)
    &=-\bs u^\top\mathbb E_n\bs\psi^\ast
      +\int_0^1(1-v)\bs u^\top
        \{-\mathbb E_n\dot{\bs\psi}_{\bs\theta^\ast+v\bs u}\}
        \bs u\,dv \\
    &\geq -\delta\|\mathbb E_n\bs\psi^\ast\|
          +\lambda\delta^2/4>0.
\end{align*}
Therefore, the minimization problem
$\min_{\|\bs\theta-\bs\theta^\ast\|\leq\delta} L_n(\bs\theta)$
has a unique minimizer satisfying
\(\|\bs\theta-\bs\theta^\ast\|<\delta\) and
\(\mathbb E_n\bs\psi_{\bs\theta}=-\nabla L_n(\bs\theta)=\bs0\). By the definition of $\widehat{\bs\theta}_{\rm lab}$, we have \(\|\widehat{\bs\theta}_{\rm lab}-\bs\theta^\ast\|<\delta\)
with probability tending to one. As this holds for every sufficiently
small fixed \(\delta>0\),
\(\|\widehat{\bs\theta}_{\rm lab}-\bs\theta^\ast\|
=o_{\mathbb P}(1)\).

Let \(\bs t=\widehat{\bs\theta}_{\rm lab}-\bs\theta^\ast\).
Expanding the sample score equation gives
\[
    \mathbb E_n\bs\psi^\ast
    =\int_0^1
      \{-\mathbb E_n\dot{\bs\psi}_{\bs\theta^\ast+v\bs t}\}
      \bs t\,dv.
\]
On \(\mathcal E_n\), \(\|v\bs t\|\leq\|\bs t\|<\delta\) for
every \(v\in[0,1]\), so
\[
    \lambda_{\min}
    (-\mathbb E_n\dot{\bs\psi}_{\bs\theta^\ast+v\bs t})
    \geq\lambda/2,
    \qquad v\in[0,1].
\]
Therefore
\[
    \frac{\lambda}{2}\|\bs t\|^2
    \leq \bs t^\top\mathbb E_n\bs\psi^\ast
    \leq \|\bs t\|\,\|\mathbb E_n\bs\psi^\ast\|.
\]
It follows that \(\|\bs t\|=O_{\mathbb P}(n^{-1/2})\).

We next show that
\[
    \sup_{0\leq v\leq1}
    \|-\mathbb E_n\dot{\bs\psi}_{\bs\theta^\ast+v\bs t}
      -\bs J\|_F
    =O_{\mathbb P}(n^{-1/2}).
\]
First,
\[
    n\E\|(\mathbb E_n-\E)\dot{\bs\psi}_{\bs\theta^\ast}\|_F^2
    \leq \E\|\dot{\bs\psi}_{\bs\theta^\ast}\|_F^2
    \leq \E(\|\widetilde{\bs X}\|^2D^2)<\infty.
\]
Using \(\bs J=-\E\dot{\bs\psi}_{\bs\theta^\ast}\), on
\(\mathcal E_n\) we have
\begin{align*}
    &\sup_{0\leq v\leq1}
    \|-\mathbb E_n\dot{\bs\psi}_{\bs\theta^\ast+v\bs t}
      -\bs J\|_F \\
    &\quad\leq
    \|(\mathbb E_n-\E)\dot{\bs\psi}_{\bs\theta^\ast}\|_F
    +\sup_{0\leq v\leq1}
      \mathbb E_n
      \|\dot{\bs\psi}_{\bs\theta^\ast+v\bs t}
        -\dot{\bs\psi}_{\bs\theta^\ast}\|_F \\
    &\quad\leq
    \|(\mathbb E_n-\E)\dot{\bs\psi}_{\bs\theta^\ast}\|_F
    +\mathbb E_n(\|\widetilde{\bs X}\|D)\|\bs t\| \\
    &\quad=
    O_{\mathbb P}(n^{-1/2})
    +O_{\mathbb P}(1)O_{\mathbb P}(n^{-1/2})
    =O_{\mathbb P}(n^{-1/2}).
\end{align*}

Using
\[
    \bs J\bs t-\mathbb E_n\bs\psi^\ast
    =
    \int_0^1
    \{\bs J+\mathbb E_n\dot{\bs\psi}_{\bs\theta^\ast+v\bs t}\}
    \bs t\,dv.
\]
we have
\begin{align*}
    \sqrt n\|\bs J\bs t-\mathbb E_n\bs\psi^\ast\|
    &\leq
    \sqrt n\|\bs t\|
    \sup_{0\leq v\leq1}
    \|-\mathbb E_n\dot{\bs\psi}_{\bs\theta^\ast+v\bs t}
      -\bs J\|_F \\
    &=O_{\mathbb P}(1)O_{\mathbb P}(n^{-1/2})
      =o_{\mathbb P}(1).
\end{align*}
Since \(\bs J\) is nonsingular, this implies
\begin{align*}
    \sqrt n(\widehat{\bs\theta}_{\rm lab}-\bs\theta^\ast) =\bs J^{-1}\sqrt n\mathbb E_n\bs\psi^\ast
      +o_{\mathbb P}(1)
\end{align*}
which proves \eqref{supp:eq:labelled-expansion}.

Finally, since
\(\widehat{\bs J}
=-\mathbb E_n\dot{\bs\psi}_{\widehat{\bs\theta}_{\rm lab}}\)
and \(\widehat{\bs\theta}_{\rm lab}=\bs\theta^\ast+\bs t\),
\begin{align*}
    \|\widehat{\bs J}-\bs J\|_F
    &=\|-\mathbb E_n\dot{\bs\psi}_{\bs\theta^\ast+\bs t}
        -\bs J\|_F \\
    &\leq
    \sup_{0\leq v\leq1}
    \|-\mathbb E_n\dot{\bs\psi}_{\bs\theta^\ast+v\bs t}
      -\bs J\|_F
    =O_{\mathbb P}(n^{-1/2}).
\end{align*}
For the fitted score, on \(\mathcal E_n\),
\[
    |\mu_{\widehat{\bs\theta}_{\rm lab}}-\mu_{\bs\theta^\ast}|
    =
    \left|\int_0^1
    (\partial_{\bs\theta}\mu_{\bs\theta^\ast+v\bs t})^\top
    \bs t\,dv\right|
    \leq D\|\bs t\|.
\]
Hence
\begin{align*}
    \|\widehat{\bs\psi}-\bs\psi^\ast\|_{n,2}
    &=
    \left\{
        \mathbb E_n\left[
            \|\widetilde{\bs X}\|^2
            |\mu_{\widehat{\bs\theta}_{\rm lab}}
             -\mu_{\bs\theta^\ast}|^2
        \right]
    \right\}^{1/2} \\
    &\leq
    \{\mathbb E_n(\|\widetilde{\bs X}\|^2D^2)\}^{1/2}
    \|\bs t\| \\
    &=O_{\mathbb P}(1)O_{\mathbb P}(n^{-1/2})
      =O_{\mathbb P}(n^{-1/2}).
\end{align*}
 Together, these bounds prove
\eqref{supp:eq:target-rates}.

\smallskip\noindent\textit{Step 2.}
We prove that
\[
    \sqrt n(\widehat{\bs\theta}_{\mathrm{san},K}-\bs\theta^\ast)
    =
    \bs J^{-1}\sqrt n
    \{\mathbb E_n(\bs\psi^\ast-\bs h_{\mathrm{san},\infty})
      +\mathbb E_N\bs h_{\mathrm{san},\infty}\}
    +o_{\mathbb P}(1).
\]

By Assumption~\ref{ass:population-score-limit},
\(\E\|\mathcal C\bs h_{\mathrm{san},K}
-\bs h_{\mathrm{san},\infty}\|^2\to0\).
Together with \eqref{supp:eq:comparator} and \(\rho_n\to\rho<1\),
this gives
\begin{align*}
    \{\E\|\mathcal C\bs h^0_K-\bs h_{\mathrm{san},\infty}\|^2\}^{1/2}
    &\leq
    \frac{1-\rho_n}{1-\rho}
    \{\E\|\mathcal C\bs h_{\mathrm{san},K}
      -\bs h_{\mathrm{san},\infty}\|^2\}^{1/2}\\
    &\quad+\frac{|\rho_n-\rho|}{1-\rho}
     \{\E\|\bs h_{\mathrm{san},\infty}\|^2\}^{1/2} \\
    &=o(1).
\end{align*}
Since \(\E\bs h_{\mathrm{san},\infty}=\bs0\) and
\(n/N\to\rho/(1-\rho)\), \eqref{supp:eq:two-sample-moment} yields
\begin{align*}
    \E\|\sqrt n\Delta_n
      (\bs h^0_K-\bs h_{\mathrm{san},\infty})\|^2
    &=(1+n/N)
      \E\|\mathcal C\bs h^0_K-\bs h_{\mathrm{san},\infty}\|^2
      =o(1),\\
    \E\|\sqrt n\Delta_n\bs h_{\mathrm{san},\infty}\|^2
    &=(1+n/N)\E\|\bs h_{\mathrm{san},\infty}\|^2
      =O(1).
\end{align*}
Hence
\(\sqrt n\Delta_n(\bs h^0_K-\bs h_{\mathrm{san},\infty})
=o_{\mathbb P}(1)\) and
\(\sqrt n\Delta_n\bs h_{\mathrm{san},\infty}
=O_{\mathbb P}(1)\).
By \eqref{supp:eq:mean-replacement},
\begin{align*}
    \sqrt n\Delta_n
    (\widehat{\bs h}_{\mathrm{san},K}-\bs h_{\mathrm{san},\infty})
    &=
    \sqrt n\Delta_n
    (\widehat{\bs h}_{\mathrm{san},K}-\bs h^0_K)
    +\sqrt n\Delta_n
    (\bs h^0_K-\bs h_{\mathrm{san},\infty}) \\
    &=o_{\mathbb P}(1).
\end{align*}
It follows that
\[
    \sqrt n\Delta_n\widehat{\bs h}_{\mathrm{san},K}
    =
    \sqrt n\Delta_n\bs h_{\mathrm{san},\infty}
    +o_{\mathbb P}(1)
    =O_{\mathbb P}(1).
\]

By \eqref{supp:eq:target-rates},
\(\|\widehat{\bs J}-\bs J\|_F=O_{\mathbb P}(n^{-1/2})\).
Since \(\bs J\) is nonsingular,
\(\|\widehat{\bs J}^{-1}\|_{\rm op}=O_{\mathbb P}(1)\), and
\begin{align*}
    \|(\widehat{\bs J}^{-1}-\bs J^{-1})
      \sqrt n\Delta_n\widehat{\bs h}_{\mathrm{san},K}\|
    &=
    \|\widehat{\bs J}^{-1}(\bs J-\widehat{\bs J})\bs J^{-1}
      \sqrt n\Delta_n\widehat{\bs h}_{\mathrm{san},K}\| \\
    &\leq
    \|\widehat{\bs J}^{-1}\|_{\rm op}
    \|\widehat{\bs J}-\bs J\|_{\rm op}
    \|\bs J^{-1}\|_{\rm op}
    \|\sqrt n\Delta_n\widehat{\bs h}_{\mathrm{san},K}\| \\
    &=O_{\mathbb P}(n^{-1/2})=o_{\mathbb P}(1).
\end{align*}

Using
\(\widehat{\bs\theta}_{\mathrm{san},K}
=\widehat{\bs\theta}_{\rm lab}
+\widehat{\bs J}^{-1}\Delta_n
 \widehat{\bs h}_{\mathrm{san},K}\)
and \eqref{supp:eq:labelled-expansion}, we obtain
\begin{align*}
    \sqrt n(\widehat{\bs\theta}_{\mathrm{san},K}-\bs\theta^\ast)
    &=
    \sqrt n(\widehat{\bs\theta}_{\rm lab}-\bs\theta^\ast)
    +\widehat{\bs J}^{-1}
     \sqrt n\Delta_n\widehat{\bs h}_{\mathrm{san},K} \\
    &=
    \bs J^{-1}\sqrt n\mathbb E_n\bs\psi^\ast
    +\{\sqrt n(\widehat{\bs\theta}_{\rm lab}-\bs\theta^\ast)
       -\bs J^{-1}\sqrt n\mathbb E_n\bs\psi^\ast\} \\
    &\quad+
    \bs J^{-1}\sqrt n\Delta_n\bs h_{\mathrm{san},\infty}
    +\bs J^{-1}\sqrt n\Delta_n
     (\widehat{\bs h}_{\mathrm{san},K}
      -\bs h_{\mathrm{san},\infty}) \\
    &\quad+
    (\widehat{\bs J}^{-1}-\bs J^{-1})
    \sqrt n\Delta_n\widehat{\bs h}_{\mathrm{san},K} \\
    &=
    \bs J^{-1}\sqrt n\mathbb E_n\bs\psi^\ast
    +o_{\mathbb P}(1)
    +\bs J^{-1}\sqrt n\Delta_n\bs h_{\mathrm{san},\infty}
    +o_{\mathbb P}(1)
    +O_{\mathbb P}(n^{-1/2}) \\
    &=
    \bs J^{-1}\sqrt n
    \{\mathbb E_n(\bs\psi^\ast-\bs h_{\mathrm{san},\infty})
      +\mathbb E_N\bs h_{\mathrm{san},\infty}\}
    +o_{\mathbb P}(1).
\end{align*}

\smallskip\noindent\textit{Step 3.}
We prove asymptotic normality and the trace inequality.

By \(\E\bs\psi^\ast=\bs0\), Step 1 and
Assumption~\ref{ass:population-score-limit},
\[
    \E(\bs\psi^\ast-\bs h_{\mathrm{san},\infty})=\bs0,
    \qquad
    \E\bs h_{\mathrm{san},\infty}=\bs0,
\]
and both vectors have finite second moments.
The central limit theorem gives
\begin{align*}
    \sqrt n\mathbb E_n
    (\bs\psi^\ast-\bs h_{\mathrm{san},\infty})
    &\rightsquigarrow
    \mathcal N\{\bs0,
      \Cov(\bs\psi^\ast-\bs h_{\mathrm{san},\infty})\},\\
    \sqrt N\mathbb E_N\bs h_{\mathrm{san},\infty}
    &\rightsquigarrow
    \mathcal N\{\bs0,\Cov(\bs h_{\mathrm{san},\infty})\}.
\end{align*}
These sample means are independent by
Assumption~\ref{ass:sampling}. Since
\(n/N\to\rho/(1-\rho)\), the expansion in Step 2 gives
\begin{align*}
    \sqrt n(\widehat{\bs\theta}_{\mathrm{san},K}-\bs\theta^\ast)
    &=
    \bs J^{-1}
    \left\{
        \sqrt n\mathbb E_n
        (\bs\psi^\ast-\bs h_{\mathrm{san},\infty})
        +\sqrt{\frac nN}\,
         \sqrt N\mathbb E_N\bs h_{\mathrm{san},\infty}
    \right\}
    +o_{\mathbb P}(1)\\
    &\rightsquigarrow
    \mathcal N\left[
        \bs0,\,
        \bs J^{-1}
        \left\{
            \Cov(\bs\psi^\ast-\bs h_{\mathrm{san},\infty})
            +\frac{\rho}{1-\rho}
             \Cov(\bs h_{\mathrm{san},\infty})
        \right\}
        \bs J^{-\top}
    \right]\\
    &=
    \mathcal N\{\bs0,\bs\Sigma(\bs h_{\mathrm{san},\infty})\}.
\end{align*}

To prove the trace inequality, write
\[
    \tr\{\bs\Sigma(\bs h)\}
    =
    \E\|\bs J^{-1}(\bs\psi^\ast-\mathcal C\bs h)\|^2
    +\frac{\rho}{1-\rho}
     \E\|\bs J^{-1}\mathcal C\bs h\|^2.
\]
Since \(\mathcal C\bs h_{\mathrm{san},\infty}
=\bs h_{\mathrm{san},\infty}\), we have
\begin{align*}
    &\left|
        \tr\{\bs\Sigma(\bs h_{\mathrm{san},K})\}
        -\tr\{\bs\Sigma(\bs h_{\mathrm{san},\infty})\}
    \right|\\
    &\quad\leq
    \left|
        \E\|\bs J^{-1}(\bs\psi^\ast
          -\mathcal C\bs h_{\mathrm{san},K})\|^2
        -\E\|\bs J^{-1}(\bs\psi^\ast
          -\bs h_{\mathrm{san},\infty})\|^2
    \right|\\
    &\qquad+
    \frac{\rho}{1-\rho}
    \left|
        \E\|\bs J^{-1}\mathcal C\bs h_{\mathrm{san},K}\|^2
        -\E\|\bs J^{-1}\bs h_{\mathrm{san},\infty}\|^2
    \right|\\
    &\quad\leq
    C\{\E\|\mathcal C\bs h_{\mathrm{san},K}
          -\bs h_{\mathrm{san},\infty}\|^2\}^{1/2}\\
    &\qquad\times
    \left[
        \{\E\|\bs\psi^\ast\|^2\}^{1/2}
        +\{\E\|\mathcal C\bs h_{\mathrm{san},K}\|^2\}^{1/2}
        +\{\E\|\bs h_{\mathrm{san},\infty}\|^2\}^{1/2}
    \right]\\
    &\quad=o(1)O(1)=o(1).
\end{align*}
The second inequality follows from
\[
\begin{aligned}
    \left|\E\|\bs u\|^2-\E\|\bs v\|^2\right|
    &=\left|\E\{(\bs u-\bs v)^\top(\bs u+\bs v)\}\right|\\
    &\leq
    \{\E\|\bs u-\bs v\|^2\}^{1/2}
    \left[
        \{\E\|\bs u\|^2\}^{1/2}
        +\{\E\|\bs v\|^2\}^{1/2}
    \right],
\end{aligned}
\]
and the last equality uses
\(\E\|\mathcal C\bs h_{\mathrm{san},K}
-\bs h_{\mathrm{san},\infty}\|^2\to0\)
from Assumption~\ref{ass:population-score-limit}.

Finally, Proposition~\ref{prop:beta-closed-form} gives
\[
    \tr\{\bs\Sigma(\bs h_{\mathrm{san},K})\}
    \leq\tr(\bs\Sigma_{\rm lab})
\]
for every \(K\). Taking limits along \(K=K_n\) yields
\[
    \tr\{\bs\Sigma(\bs h_{\mathrm{san},\infty})\}
    =
    \lim_{n\to\infty}
    \tr\{\bs\Sigma(\bs h_{\mathrm{san},K_n})\}
    \leq\tr(\bs\Sigma_{\rm lab}).
\]
\end{proof}

\subsection{Proof of Corollary~\ref{cor:variance-consistency}}

\begin{proof}
\smallskip\noindent\textit{Step 1.}
We prove that
\begin{align*}
    \widehat{\bs V}_n
    (\widehat{\bs\psi}-\widehat{\bs h}_{\mathrm{san},K})
    &\xrightarrow{\mathbb P}
    \Cov(\bs\psi^\ast-\bs h_{\mathrm{san},\infty}),\\
    \widehat{\bs V}_N(\widehat{\bs h}_{\mathrm{san},K})
    &\xrightarrow{\mathbb P}
    \Cov(\bs h_{\mathrm{san},\infty}).
\end{align*}

Step 2 of the proof of Theorem~\ref{thm:san-asymptotic}
establishes
\(\E\|\mathcal C\bs h_K^0-\bs h_{\mathrm{san},\infty}\|^2=o(1)\).
For \(m\in\{n,N\}\),
\[
    \E\left[
        \mathbb E_m
        \|\mathcal C\bs h_K^0-\bs h_{\mathrm{san},\infty}\|^2
    \right]
    =
    \E\|\mathcal C\bs h_K^0-\bs h_{\mathrm{san},\infty}\|^2
    =o(1).
\]
Markov's inequality therefore gives
\[
    \mathbb E_m
    \|\mathcal C\bs h_K^0-\bs h_{\mathrm{san},\infty}\|^2
    =o_{\mathbb P}(1).
\]
Recall that
\(\bs\delta_K=\widehat{\bs h}_{\mathrm{san},K}-\bs h_K^0\).
Then
\[
    \mathcal C_m
    (\widehat{\bs h}_{\mathrm{san},K}-\bs h_{\mathrm{san},\infty})
    =
    \mathcal C_m\bs\delta_K
    +\mathcal C_m
     (\mathcal C\bs h_K^0-\bs h_{\mathrm{san},\infty}).
\]
Since
\(\mathbb E_m\|\mathcal C_m\bs a\|^2
\leq\mathbb E_m\|\bs a\|^2\),
\eqref{supp:eq:variance-replacement} gives
\begin{align*}
    &\mathbb E_m
    \|\mathcal C_m
      (\widehat{\bs h}_{\mathrm{san},K}
       -\bs h_{\mathrm{san},\infty})\|^2\\
    &\quad\leq
    2\mathbb E_m\|\mathcal C_m\bs\delta_K\|^2
    +2\mathbb E_m
      \|\mathcal C\bs h_K^0-\bs h_{\mathrm{san},\infty}\|^2\\
    &\quad=
    2\tr\{\widehat{\bs V}_m(\bs\delta_K)\}
    +o_{\mathbb P}(1)
    =o_{\mathbb P}(1).
\end{align*}
Similarly, \eqref{supp:eq:target-rates} yields
\begin{align*}
    &\mathbb E_n
    \left\|
        \mathcal C_n
        \{(\widehat{\bs\psi}-\widehat{\bs h}_{\mathrm{san},K})
          -(\bs\psi^\ast-\bs h_{\mathrm{san},\infty})\}
    \right\|^2\\
    &\quad\leq
    2\mathbb E_n\|\widehat{\bs\psi}-\bs\psi^\ast\|^2
    +2\mathbb E_n
      \|\mathcal C_n
        (\widehat{\bs h}_{\mathrm{san},K}
         -\bs h_{\mathrm{san},\infty})\|^2\\
    &\quad=O_{\mathbb P}(n^{-1})+o_{\mathbb P}(1)
    =o_{\mathbb P}(1).
\end{align*}

For any fixed vector function \(\bs v\) with
\(\E\|\bs v\|^2<\infty\), the law of large numbers gives
\begin{align*}
    \mathbb E_m\|\mathcal C_m\bs v\|^2
    &=
    \mathbb E_m\|\bs v\|^2-\|\mathbb E_m\bs v\|^2
    \xrightarrow{\mathbb P}
    \E\|\bs v\|^2-\|\E\bs v\|^2,\\
    \widehat{\bs V}_m(\bs v)
    &=
    \mathbb E_m(\bs v\bs v^\top)
    -(\mathbb E_m\bs v)(\mathbb E_m\bs v)^\top
    \xrightarrow{\mathbb P}\Cov(\bs v).
\end{align*}
For any sample vectors \(\bs u,\bs v\),
\begin{align*}
    \widehat{\bs V}_m(\bs u)-\widehat{\bs V}_m(\bs v)
    &=
    \mathbb E_m\left[
        \{\mathcal C_m\bs v+\mathcal C_m(\bs u-\bs v)\}
        \{\mathcal C_m\bs v+\mathcal C_m(\bs u-\bs v)\}^\top
        -(\mathcal C_m\bs v)(\mathcal C_m\bs v)^\top
    \right]\\
    &=
    \mathbb E_m[
        \{\mathcal C_m(\bs u-\bs v)\}(\mathcal C_m\bs v)^\top
    ]\\
    &\quad+
    \mathbb E_m[
        (\mathcal C_m\bs v)\{\mathcal C_m(\bs u-\bs v)\}^\top
    ]\\
    &\quad+
    \mathbb E_m\left[
        \{\mathcal C_m(\bs u-\bs v)\}
        \{\mathcal C_m(\bs u-\bs v)\}^\top
    \right].
\end{align*}
Consequently,
\begin{align*}
    \|\widehat{\bs V}_m(\bs u)-\widehat{\bs V}_m(\bs v)\|_F
    &\leq
    2\{\mathbb E_m\|\mathcal C_m(\bs u-\bs v)\|^2\}^{1/2}
     \{\mathbb E_m\|\mathcal C_m\bs v\|^2\}^{1/2}\\
    &\quad+
    \mathbb E_m\|\mathcal C_m(\bs u-\bs v)\|^2.
\end{align*}
Applying this inequality to the two pairs of scores gives
\begin{align*}
    &\|\widehat{\bs V}_n
      (\widehat{\bs\psi}-\widehat{\bs h}_{\mathrm{san},K})
      -\widehat{\bs V}_n
      (\bs\psi^\ast-\bs h_{\mathrm{san},\infty})\|_F\\
    &\quad\leq
    2\left[
        \mathbb E_n
        \left\|\mathcal C_n
        \{(\widehat{\bs\psi}-\widehat{\bs h}_{\mathrm{san},K})
          -(\bs\psi^\ast-\bs h_{\mathrm{san},\infty})\}
        \right\|^2
    \right]^{1/2}\\
    &\qquad\times
    \{\mathbb E_n
      \|\mathcal C_n(\bs\psi^\ast-\bs h_{\mathrm{san},\infty})\|^2
    \}^{1/2}
    +o_{\mathbb P}(1)\\
    &\quad=o_{\mathbb P}(1)O_{\mathbb P}(1)+o_{\mathbb P}(1)
    =o_{\mathbb P}(1),\\
    &\|\widehat{\bs V}_N(\widehat{\bs h}_{\mathrm{san},K})
      -\widehat{\bs V}_N(\bs h_{\mathrm{san},\infty})\|_F\\
    &\quad\leq
    2\{\mathbb E_N
      \|\mathcal C_N(\widehat{\bs h}_{\mathrm{san},K}
                    -\bs h_{\mathrm{san},\infty})\|^2\}^{1/2}
    \{\mathbb E_N\|\mathcal C_N\bs h_{\mathrm{san},\infty}\|^2\}^{1/2}
    +o_{\mathbb P}(1)\\
    &\quad=o_{\mathbb P}(1)O_{\mathbb P}(1)+o_{\mathbb P}(1)
    =o_{\mathbb P}(1).
\end{align*}
Together with the law of large numbers, these bounds imply
\begin{align*}
    \widehat{\bs V}_n
    (\widehat{\bs\psi}-\widehat{\bs h}_{\mathrm{san},K})
    &=
    \widehat{\bs V}_n
    (\bs\psi^\ast-\bs h_{\mathrm{san},\infty})
    +o_{\mathbb P}(1)\\
    &=
    \Cov(\bs\psi^\ast-\bs h_{\mathrm{san},\infty})
    +o_{\mathbb P}(1),\\
    \widehat{\bs V}_N(\widehat{\bs h}_{\mathrm{san},K})
    &=
    \widehat{\bs V}_N(\bs h_{\mathrm{san},\infty})
    +o_{\mathbb P}(1)\\
    &=
    \Cov(\bs h_{\mathrm{san},\infty})
    +o_{\mathbb P}(1).
\end{align*}

\smallskip\noindent\textit{Step 2.}
We prove consistency of
\(\widehat{\bs\Sigma}_{\mathrm{san},K}\)
and the limiting coverage of the Wald interval.
Recall that
\[
    \bs\Omega(\bs h_{\mathrm{san},\infty})
    =
    \Cov(\bs\psi^\ast-\bs h_{\mathrm{san},\infty})
    +\frac{\rho}{1-\rho}\Cov(\bs h_{\mathrm{san},\infty}).
\]
By Step 1 and \(n/N\to\rho/(1-\rho)\),
\begin{align*}
    \|\widehat{\bs\Omega}_{\mathrm{san},K}
      -\bs\Omega(\bs h_{\mathrm{san},\infty})\|_{\rm op}
    &\leq
    \|\widehat{\bs V}_n
      (\widehat{\bs\psi}-\widehat{\bs h}_{\mathrm{san},K})
      -\Cov(\bs\psi^\ast-\bs h_{\mathrm{san},\infty})\|_{\rm op}\\
    &\quad+
    \frac nN
    \|\widehat{\bs V}_N(\widehat{\bs h}_{\mathrm{san},K})
      -\Cov(\bs h_{\mathrm{san},\infty})\|_{\rm op}\\
    &\quad+
    \left|\frac nN-\frac{\rho}{1-\rho}\right|
    \|\Cov(\bs h_{\mathrm{san},\infty})\|_{\rm op}\\
    &=o_{\mathbb P}(1)+O(1)o_{\mathbb P}(1)+o(1)O(1)
    =o_{\mathbb P}(1).
\end{align*}
Also, \eqref{supp:eq:target-rates} and nonsingularity of \(\bs J\)
give
\[
    \|\widehat{\bs J}^{-1}-\bs J^{-1}\|_{\rm op}
    \leq
    \|\widehat{\bs J}^{-1}\|_{\rm op}
    \|\widehat{\bs J}-\bs J\|_{\rm op}
    \|\bs J^{-1}\|_{\rm op}
    =O_{\mathbb P}(n^{-1/2}).
\]
Thus
\begin{align*}
    &\widehat{\bs J}^{-1}\widehat{\bs\Omega}_{\mathrm{san},K}
     \widehat{\bs J}^{-\top}
     -\bs J^{-1}\bs\Omega(\bs h_{\mathrm{san},\infty})\bs J^{-\top}\\
    &\quad=
    (\widehat{\bs J}^{-1}-\bs J^{-1})
    \widehat{\bs\Omega}_{\mathrm{san},K}\widehat{\bs J}^{-\top}\\
    &\qquad+
    \bs J^{-1}
    \{\widehat{\bs\Omega}_{\mathrm{san},K}
      -\bs\Omega(\bs h_{\mathrm{san},\infty})\}
    \widehat{\bs J}^{-\top}\\
    &\qquad+
    \bs J^{-1}\bs\Omega(\bs h_{\mathrm{san},\infty})
    (\widehat{\bs J}^{-\top}-\bs J^{-\top})\\
    &\quad=
    O_{\mathbb P}(n^{-1/2})
    +o_{\mathbb P}(1)
    +O_{\mathbb P}(n^{-1/2})
    =o_{\mathbb P}(1).
\end{align*}

Since \(p\) is fixed and \(K^5/n\to0\),
\[
    \frac{d_{\rm eff}}n
    =\frac{2p+K+2}{n}\longrightarrow0,
    \qquad
    \frac{n}{n-d_{\rm eff}}-1
    =\frac{d_{\rm eff}/n}{1-d_{\rm eff}/n}
    \longrightarrow0.
\]
Using
\(\bs\Sigma(\bs h_{\mathrm{san},\infty})
=\bs J^{-1}\bs\Omega(\bs h_{\mathrm{san},\infty})\bs J^{-\top}\),
we conclude that
\begin{align*}
    &\widehat{\bs\Sigma}_{\mathrm{san},K}
      -\bs\Sigma(\bs h_{\mathrm{san},\infty})\\
    &\quad=
    \left(\frac{n}{n-d_{\rm eff}}-1\right)
    \widehat{\bs J}^{-1}\widehat{\bs\Omega}_{\mathrm{san},K}
    \widehat{\bs J}^{-\top}\\
    &\qquad+
    \{\widehat{\bs J}^{-1}\widehat{\bs\Omega}_{\mathrm{san},K}
      \widehat{\bs J}^{-\top}
      -\bs J^{-1}\bs\Omega(\bs h_{\mathrm{san},\infty})
       \bs J^{-\top}\}\\
    &\quad=o(1)O_{\mathbb P}(1)+o_{\mathbb P}(1)
    =o_{\mathbb P}(1).
\end{align*}

For any fixed nonzero \(\bs c\in\mathbb R^p\),
Assumption~\ref{ass:population-score-limit} gives
\[
    \bs c^\top\widehat{\bs\Sigma}_{\mathrm{san},K}\bs c
    \xrightarrow{\mathbb P}
    \bs c^\top\bs\Sigma(\bs h_{\mathrm{san},\infty})\bs c>0.
\]
Theorem~\ref{thm:san-asymptotic} and Slutsky's theorem yield
\[
    \frac{
        \sqrt n\,\bs c^\top
        (\widehat{\bs\theta}_{\mathrm{san},K}-\bs\theta^\ast)
    }{
        (\bs c^\top\widehat{\bs\Sigma}_{\mathrm{san},K}\bs c)^{1/2}
    }
    \rightsquigarrow\mathcal N(0,1).
\]
Therefore,
\[
    \mathbb P\left\{
        \frac{
            \sqrt n\,|\bs c^\top
            (\widehat{\bs\theta}_{\mathrm{san},K}-\bs\theta^\ast)|
        }{
            (\bs c^\top\widehat{\bs\Sigma}_{\mathrm{san},K}\bs c)^{1/2}
        }
        \leq z_{1-\alpha/2}
    \right\}
    \longrightarrow1-\alpha.
\]
\end{proof}

\subsection{Proof of Proposition~\ref{thm:conditional-mean-preservation}}

\begin{proof}
 It suffices to prove
\(\E(Y\mid S)=g(T)\) almost surely.
The conclusion holds trivially if \(g(T)\) is a constant almost surely.
We now consider the case where \(g(T)\) is nonconstant.

Write \(\bs\Sigma_Q=\Cov(\bs Q)\).
For \(\bs v\ne\bs0\) and \(U=v_0+\bs v^\top\bs Q\),
ellipticity implies the following identity \citep{cambanis1981elliptical}
\[
    \E(\bs Q-\E\bs Q\mid U)
    =
    \frac{\bs\Sigma_Q\bs v}
         {\bs v^\top\bs\Sigma_Q\bs v}
    (U-\E U).
\]
Consequently, for any measurable \(r\) satisfying
\(\E[(1+\|\bs Q\|)|r(U)|]<\infty\),
\begin{equation}
\begin{aligned}
    \Cov\{\bs Q,r(U)\}
    &=
    \E[\E(\bs Q-\E\bs Q\mid U)r(U)]\\
    &=
    \frac{\bs\Sigma_Q\bs v}
         {\bs v^\top\bs\Sigma_Q\bs v}
    \E\{(U-\E U)r(U)\}\\
    &=
    \frac{\bs\Sigma_Q\bs v}
         {\bs v^\top\bs\Sigma_Q\bs v}
    \Cov\{U,r(U)\}.
\end{aligned}
\label{supp:eq:elliptical-identity}
\end{equation}

Write
\(\bs\gamma^\ast=(\gamma_0^\ast,\bs v^{\ast\top})^\top\)
and
\(\eta=\gamma_0^\ast+\bs v^{\ast\top}\bs Q\),
so \(S=A'(\eta)\).
Since
\[
    \E\left[
        \begin{pmatrix}1\\ \bs Q\end{pmatrix}
        \{Y- S\}
    \right]=\bs0,
\]
it follows that
\begin{align*}
    \Cov\{\bs Q,A'(\eta)\}
    &=
    \E(\bs Q S)-\E\bs Q\,\E S\\
    &=
    \E(\bs Q Y)-\E\bs Q\,\E Y\\
    &=
    \E[\bs Q\,\E(Y\mid\bs Q)]
    -\E\bs Q\,\E\{\E(Y\mid\bs Q)\}\\
    &=
    \E\{\bs Qg(T)\}-\E\bs Q\,\E g(T)
    =\Cov\{\bs Q,g(T)\}.
\end{align*}
Since \(g(T)=\E(Y\mid\bs Q)\), conditional Jensen's inequality gives
\[
\begin{aligned}
    \E[(1+\|\bs Q\|)|g(T)|]
    &\leq
    \E[(1+\|\bs Q\|)\E(|Y|\mid\bs Q)]\\
    &=\E[(1+\|\bs Q\|)|Y|] <\infty,
\end{aligned}
\]
where the last inequality follows from
\(\E[\|\widetilde{\bs Q}\|(|Y|+|S|)]<\infty\)
in the proposition.
Applying \eqref{supp:eq:elliptical-identity} with
\(U=T=\bs\alpha^\top\bs Q\) and \(r=g\), we obtain
\[
    \Cov\{\bs Q,g(T)\}
    =
    \frac{\Cov\{T,g(T)\}}
         {\bs\alpha^\top\bs\Sigma_Q\bs\alpha}
    \bs\Sigma_Q\bs\alpha
    \ne\bs0,
\]
where the last inequality follows from
\(\Cov\{T,g(T)\}\ne0\) and nonsingularity of \(\bs\Sigma_Q\).
Hence \(\bs v^\ast\ne\bs0\), since \(\bs v^\ast=\bs0\) would
make \(A'(\eta)\) constant and
\(\Cov\{\bs Q,A'(\eta)\}=\bs0\).

For an independent copy \(\eta'\) of \(\eta\),
\[
    \Var(\eta)=\bs v^{\ast\top}\bs\Sigma_Q\bs v^\ast>0,
\]
and strict monotonicity of \(A'\) gives
\[
    \Cov\{\eta,A'(\eta)\}
    =
    \frac12\E[
        (\eta-\eta')\{A'(\eta)-A'(\eta')\}
    ]>0.
\]
Applying \eqref{supp:eq:elliptical-identity} with
\(U=\eta\) and \(r=A'\) therefore gives
\[
    \frac{\Cov\{\eta,A'(\eta)\}}
         {\bs v^{\ast\top}\bs\Sigma_Q\bs v^\ast}
    \bs\Sigma_Q\bs v^\ast
    =
    \Cov\{\bs Q,A'(\eta)\}
    =
    \Cov\{\bs Q,g(T)\}
    =
    \frac{\Cov\{T,g(T)\}}
         {\bs\alpha^\top\bs\Sigma_Q\bs\alpha}
    \bs\Sigma_Q\bs\alpha.
\]
Therefore
\[
    \bs v^\ast=c\bs\alpha,
    \qquad
    c=
    \frac{\Cov\{T,g(T)\}}
         {\bs\alpha^\top\bs\Sigma_Q\bs\alpha}
    \frac{\bs v^{\ast\top}\bs\Sigma_Q\bs v^\ast}
         {\Cov\{\eta,A'(\eta)\}}
    \ne0.
\]
Thus
\[
    \eta=\gamma_0^\ast+cT,
    \qquad
    S=A'(\gamma_0^\ast+cT),
    \qquad
    T=\frac{(A')^{-1}(S)-\gamma_0^\ast}{c}.
\]
In particular, \(T\) is measurable with respect to \(S\), so
\[
    \E(Y\mid S)
    =
    \E\{\E(Y\mid\bs Q)\mid S\}
    =
    \E\{g(T)\mid S\}
    =
    g(T)
    =
    \E(Y\mid f,\bs X).
\]
\end{proof}

\subsection{Gaussian single-index example}
\label{supp:gaussian-example}

The following proposition verifies the single-index assumption
\eqref{eq:conditional-single-index} for the Gaussian model
in Remark~\ref{rem:gaussian-single-index} of the main paper.
\begin{proposition}
\label{supp:prop:gaussian-single-index}
Let \(\bs X\sim N(\bs0,\Sigma_X)\) and
\(\bs W\sim N(\bs0,\Sigma_W)\) be independent, with
\(\Sigma_X\) positive definite. Suppose that \(\E|Y|<\infty\) and
\[
    \E(Y\mid\bs X,\bs W)
    =A'(\bs\delta_X^\top\bs X+\bs\delta_W^\top\bs W),
    \qquad
    \tau(f)=\widetilde{\bs\delta}_X^\top\bs X
           +\widetilde{\bs\delta}_W^\top\bs W,
\]
where \(\tau\) has a measurable inverse.
For \(\bs Q=(\tau(f),\bs X^\top)^\top\), assume that
\[
    v_f=\widetilde{\bs\delta}_W^\top\Sigma_W
        \widetilde{\bs\delta}_W>0,
\]
and define
\[
    a=\frac{\bs\delta_W^\top\Sigma_W
        \widetilde{\bs\delta}_W}{v_f}.
\]
Then \(\bs Q\) is multivariate normal with a positive definite
covariance matrix, and
\[
    \E(Y\mid f,\bs X)=g(\bs\alpha^\top\bs Q)
    \quad\text{almost surely},
    \qquad
    \bs\alpha=
    \begin{pmatrix}
        a\\
        \bs\delta_X-a\widetilde{\bs\delta}_X
    \end{pmatrix},
\]
where
\[
    g(t)=\E\{A'(t+\varepsilon)\},
    \qquad
    \varepsilon\sim
    N\!\left(0,\bs\delta_W^\top\Sigma_W\bs\delta_W-a^2v_f\right).
\]
\end{proposition}

\begin{proof}
For any scalar \(c\) and vector \(\bs d\) of the same dimension
as \(\bs X\), independence of \(\bs X\) and \(\bs W\) gives
\[
\begin{aligned}
    \Var\{c\tau(f)+\bs d^\top\bs X\}
    &=(c\widetilde{\bs\delta}_X+\bs d)^\top
      \Sigma_X(c\widetilde{\bs\delta}_X+\bs d)+c^2v_f.
\end{aligned}
\]
If \(c\ne0\), the last term is positive since \(v_f>0\).
If \(c=0\) and \(\bs d\ne\bs0\), the first term is positive
since \(\Sigma_X\) is positive definite.
Thus \(\Cov(\bs Q)\) is positive definite. Moreover, \(\bs Q\)
is a linear transformation of the jointly Gaussian vector
\((\bs X^\top,\bs W^\top)^\top\), so it is multivariate normal.

Put
\[
    U=\widetilde{\bs\delta}_W^\top\bs W,
    \qquad
    \varepsilon=\bs\delta_W^\top\bs W-aU.
\]
By the definition of \(a\),
\[
    \Cov(\varepsilon,U)
    =\bs\delta_W^\top\Sigma_W\widetilde{\bs\delta}_W-av_f=0.
\]
Independence of \(\bs X\) and \(\bs W\) also gives
\(\Cov(\varepsilon,\bs X)=\bs0\).
Since \((\varepsilon,U,\bs X^\top)^\top\) is jointly Gaussian,
\(\varepsilon\) is independent of \((U,\bs X)\), and hence of
\(\bs Q\). Moreover,
\[
\begin{aligned}
    \E\varepsilon&=0,\\
    \Var(\varepsilon)
    &=\bs\delta_W^\top\Sigma_W\bs\delta_W
      -2a\bs\delta_W^\top\Sigma_W\widetilde{\bs\delta}_W
      +a^2v_f\\
    &=\bs\delta_W^\top\Sigma_W\bs\delta_W-a^2v_f.
\end{aligned}
\]
Using \(U=\tau(f)-\widetilde{\bs\delta}_X^\top\bs X\), we obtain
\[
\begin{aligned}
    \bs\delta_X^\top\bs X+\bs\delta_W^\top\bs W
    &=\bs\delta_X^\top\bs X+aU+\varepsilon\\
    &=a\tau(f)
      +(\bs\delta_X-a\widetilde{\bs\delta}_X)^\top\bs X
      +\varepsilon\\
    &=\bs\alpha^\top\bs Q+\varepsilon.
\end{aligned}
\]
The assumption \(\E|Y|<\infty\) implies
\[
    \E|A'(\bs\alpha^\top\bs Q+\varepsilon)|
    =\E|\E(Y\mid\bs X,\bs W)|
    \leq\E|Y|<\infty,
\]
so \(g(\bs\alpha^\top\bs Q)\) is finite almost surely.
Since \(\tau\) has a measurable inverse,
\(\sigma(f,\bs X)=\sigma(\bs Q)\). Therefore,
\[
\begin{aligned}
    \E(Y\mid f,\bs X)
    &=\E\{\E(Y\mid\bs X,\bs W)\mid\bs Q\}\\
    &=\E\{A'(\bs\alpha^\top\bs Q+\varepsilon)\mid\bs Q\}\\
    &=g(\bs\alpha^\top\bs Q),
\end{aligned}
\]
where the last equality follows from independence of
\(\varepsilon\) and \(\bs Q\).
\end{proof}

\subsection{Proofs of the technical lemmas}
\label{supp:technical-proofs}

\begin{proof}[Proof of Lemma~\ref{supp:lem:normal-equations}]
Write
\[
    \bs\eta=(\bs\theta^\top,\bs\gamma^\top)^\top,
    \qquad
    \bs\eta^\ast
    =(\bs\theta^{\ast\top},\bs\gamma^{\ast\top})^\top,
    \qquad
    \widehat{\bs\eta}
    =(\widehat{\bs\theta}_{\rm lab}^\top,
      \widehat{\bs\gamma}^\top)^\top.
\]
By \eqref{supp:eq:labelled-expansion} and
Assumption~\ref{ass:score-learning}(i),
\[
    \|\widehat{\bs\eta}-\bs\eta^\ast\|
    =O_{\mathbb P}(n^{-1/2}).
\]
Define
\[
    \bs B_{\bs\eta,K}
    =\widetilde{\bs X}
      [\mu_{\bs\theta},\bs b_K(S_{\bs\gamma})^\top],
    \qquad
    \bs Z_{\bs\eta,K}
    =\bs J^{-1}\mathcal C\bs B_{\bs\eta,K}\bs G_K^{-1/2},
\]
and write
\(\bs Z_K=\bs Z_{\bs\eta^\ast,K}\) and
\(\widehat{\bs Z}_K=\bs Z_{\widehat{\bs\eta},K}\).
Derivatives of matrix-valued functions are taken entrywise.
Since \(\bs\eta\in\mathbb R^{2p+1}\), we use
\[
\begin{aligned}
    \|\partial_{\bs\eta}\bs Z_{\bs\eta,K}\|_F^2
    &:=\sum_{j=1}^{2p+1}
       \left\|\frac{\partial\bs Z_{\bs\eta,K}}{\partial\eta_j}\right\|_F^2,\\
    (\partial_{\bs\eta}\bs Z_{\bs\eta,K})\bs v
    &:=\sum_{j=1}^{2p+1}v_j
       \frac{\partial\bs Z_{\bs\eta,K}}{\partial\eta_j},
       \qquad \bs v\in\mathbb R^{2p+1}.
\end{aligned}
\]
The same conventions apply to \(\bs B_{\bs\eta,K}\) and to differences
of these derivatives. Expectations and centering operators act entrywise.
All suprema below are over fixed convex neighborhoods of
\(\bs\theta^\ast\) and \(\bs\gamma^\ast\) contained in
\(\mathcal U_\theta\) and \(\mathcal U_\gamma\), respectively.

\smallskip\noindent\textit{Step 1: Bounds for \(\bs Z_{\bs\eta,K}\).}
By the definition of \(\bs G_K\),
\begin{align*}
    \E\bs Z_K&=\bs0,\\
    \E(\bs Z_K^\top\bs Z_K)
    &=
    \bs G_K^{-1/2}
    \E\{(\mathcal C\bs B_{\bs\eta^\ast,K})^\top
        \bs J^{-\top}\bs J^{-1}
        \mathcal C\bs B_{\bs\eta^\ast,K}\}
    \bs G_K^{-1/2}\\
    &=\bs G_K^{-1/2}\bs G_K\bs G_K^{-1/2}
    =\bs I_{K+1}.
\end{align*}
Consequently,
\begin{equation}
    \E\bs Z_K=\bs0,
    \qquad
    \E(\bs Z_K^\top\bs Z_K)=\bs I_{K+1},
    \qquad
    \E\|\bs Z_K\|_F^2=\tr(\bs I_{K+1})=K+1.
    \label{supp:eq:normalized-design}
\end{equation}

Since
\[
    \partial_{\bs\gamma}\bs b_K(S_{\bs\gamma})
    =
    \bs b_K'(S_{\bs\gamma})
    (\partial_{\bs\gamma}S_{\bs\gamma})^\top,
\]
\eqref{supp:eq:spline-bounds} and
Assumption~\ref{ass:score-learning}(ii) give
\begin{align*}
    \sup_{\bs\gamma}
    \|\partial_{\bs\gamma}\bs b_K(S_{\bs\gamma})\|_F
    &=
    \sup_{\bs\gamma}
    \{\|\bs b_K'(S_{\bs\gamma})\|
      \|\partial_{\bs\gamma}S_{\bs\gamma}\|\}\\
    &\leq CKD^{1/2},\\
    \sup_{\bs\eta}\|\bs B_{\bs\eta,K}\|_F
    &=
    \|\widetilde{\bs X}\|
    \sup_{\bs\eta}
    \{\mu_{\bs\theta}^2
      +\|\bs b_K(S_{\bs\gamma})\|^2\}^{1/2}\\
    &\leq
    \|\widetilde{\bs X}\|(D^2+1)^{1/2}
    \leq C\|\widetilde{\bs X}\|D,\\
    \sup_{\bs\eta}
    \|\partial_{\bs\eta}\bs B_{\bs\eta,K}\|_F
    &=
    \|\widetilde{\bs X}\|
    \sup_{\bs\eta}
    \{\|\partial_{\bs\theta}\mu_{\bs\theta}\|^2
      +\|\partial_{\bs\gamma}
          \bs b_K(S_{\bs\gamma})\|_F^2\}^{1/2}\\
    &\leq
    \|\widetilde{\bs X}\|(D^2+CK^2D)^{1/2}
    \leq CK\|\widetilde{\bs X}\|D.
\end{align*}
Put
\(F=\|\widetilde{\bs X}\|D+\E(\|\widetilde{\bs X}\|D)\).
Then
\begin{align*}
    \sup_{\bs\eta}
    \|\mathcal C\bs B_{\bs\eta,K}\|_F
    &\leq
    \sup_{\bs\eta}\|\bs B_{\bs\eta,K}\|_F
    +\E\sup_{\bs\eta}\|\bs B_{\bs\eta,K}\|_F
    \leq CF,\\
    \sup_{\bs\eta}
    \|\mathcal C\partial_{\bs\eta}\bs B_{\bs\eta,K}\|_F
    &\leq
    \sup_{\bs\eta}
    \|\partial_{\bs\eta}\bs B_{\bs\eta,K}\|_F
    +\E\sup_{\bs\eta}
    \|\partial_{\bs\eta}\bs B_{\bs\eta,K}\|_F
    \leq CKF.
\end{align*}
By Assumption~\ref{ass:score-learning}(ii),
\[
    \E F^4
    \leq
    8\E(\|\widetilde{\bs X}\|^4D^4)
    +8\{\E(\|\widetilde{\bs X}\|D)\}^4
    \leq16\E(\|\widetilde{\bs X}\|^4D^4)<\infty.
\]
The integrable derivative bound also gives
\[
    \partial_{\bs\eta}\mathcal C\bs B_{\bs\eta,K}
    =
    \mathcal C\partial_{\bs\eta}\bs B_{\bs\eta,K}.
\]
Since Assumption~\ref{ass:score-learning}(iv) implies
\[
    \|\bs G_K^{-1/2}\|_{\rm op}
    =\lambda_{\min}(\bs G_K)^{-1/2}
    \leq c_G^{-1/2}K^{1/2},
\]
we obtain
\begin{equation}
\begin{aligned}
    \sup_{\bs\eta}\|\bs Z_{\bs\eta,K}\|_F
    &\leq
    \|\bs J^{-1}\|_{\rm op}
    \|\bs G_K^{-1/2}\|_{\rm op}\,CF
    \leq CK^{1/2}F,\\
    \sup_{\bs\eta}
    \|\partial_{\bs\eta}\bs Z_{\bs\eta,K}\|_F
    &\leq
    \|\bs J^{-1}\|_{\rm op}
    \|\bs G_K^{-1/2}\|_{\rm op}\,CKF
    \leq CK^{3/2}F.
\end{aligned}
\label{supp:eq:design-envelopes}
\end{equation}

\smallskip\noindent
\textit{Step 2: Bounds for
\(\mathbb E_m(\bs Z_K^\top\bs Z_K)-\bs I_{K+1}\),
\(\mathbb E_m\bs Z_K\), and
\(\mathbb E_m\|\widehat{\bs Z}_K-\bs Z_K\|_F^2\).}

For \(m\in\{n,N\}\), independence of the observations and
\eqref{supp:eq:normalized-design} give
\begin{align*}
    \E\|\mathbb E_m(\bs Z_K^\top\bs Z_K)-\bs I_{K+1}\|_F^2
    &=
    \frac1m
    \{\E\|\bs Z_K^\top\bs Z_K\|_F^2-\|\bs I_{K+1}\|_F^2\}\\
    &\leq\frac1m\E\|\bs Z_K\|_F^4
    \leq\frac{CK^2}{m}\E F^4,\\
    \E\|\mathbb E_m\bs Z_K\|_F^2
    &=\frac1m\E\|\bs Z_K\|_F^2
    \leq\frac{K+1}{m}.
\end{align*}
Markov's inequality and
\(\|\cdot\|_{\rm op}\leq\|\cdot\|_F\) therefore imply
\begin{equation}
    \|\mathbb E_m(\bs Z_K^\top\bs Z_K)-\bs I_{K+1}\|_{\rm op}
    =O_{\mathbb P}(K/\sqrt m),
    \qquad
    \|\mathbb E_m\bs Z_K\|_F
    =O_{\mathbb P}(\sqrt{K/m}).
    \label{supp:eq:fixed-design-moments}
\end{equation}

Choose a fixed \(\delta>0\) such that
\[
    \{\bs\eta:\|\bs\eta-\bs\eta^\ast\|\leq\delta\}
    \subseteq\mathcal U_\theta\times\mathcal U_\gamma.
\]
The bounds in \eqref{supp:eq:design-envelopes} then hold
for all \(\|\bs\eta-\bs\eta^\ast\|\leq\delta\).
Since
\(\|\widehat{\bs\eta}-\bs\eta^\ast\|
=O_{\mathbb P}(n^{-1/2})\),
\[
    \mathbb P\{
        \|\widehat{\bs\eta}-\bs\eta^\ast\|\leq\delta
    \}\longrightarrow1.
\]
On this event,
\[
    \|\bs\eta^\ast
      +u(\widehat{\bs\eta}-\bs\eta^\ast)-\bs\eta^\ast\|
    =u\|\widehat{\bs\eta}-\bs\eta^\ast\|
    \leq\delta,
    \qquad 0\leq u\leq1.
\]
Thus
\begin{align*}
    \widehat{\bs Z}_K-\bs Z_K
    &=
    \int_0^1
    \partial_{\bs\eta}
    \bs Z_{\bs\eta^\ast+u(\widehat{\bs\eta}-\bs\eta^\ast),K}
    (\widehat{\bs\eta}-\bs\eta^\ast)\,du,\\
    \|\widehat{\bs Z}_K-\bs Z_K\|_F
    &\leq
    \int_0^1
    \|\partial_{\bs\eta}
      \bs Z_{\bs\eta^\ast+u(\widehat{\bs\eta}-\bs\eta^\ast),K}\|_F
    \,du\,
    \|\widehat{\bs\eta}-\bs\eta^\ast\|\\
    &\leq CK^{3/2}F
    \|\widehat{\bs\eta}-\bs\eta^\ast\|.
\end{align*}
Since \(\E F^4<\infty\), the law of large numbers gives
\(\mathbb E_mF^2\xrightarrow{\mathbb P}\E F^2<\infty\).
Consequently,
\begin{equation}
\begin{aligned}
    \{\mathbb E_m
       \|\widehat{\bs Z}_K-\bs Z_K\|_F^2\}^{1/2}
    &\leq
    CK^{3/2}(\mathbb E_mF^2)^{1/2}
    \|\widehat{\bs\eta}-\bs\eta^\ast\|\\
    &=K^{3/2}O_{\mathbb P}(1)O_{\mathbb P}(n^{-1/2})
    =O_{\mathbb P}\{(K^3/n)^{1/2}\}.
\end{aligned}
\label{supp:eq:design-perturbation}
\end{equation}

\smallskip\noindent
\textit{Step 3: Bounding
\(\|\bs G_K^{-1/2}\widehat{\bs G}_K\bs G_K^{-1/2}-\bs I_{K+1}\|_{\rm op}\)
and proving invertibility of \(\widehat{\bs G}_K\).}
Write
\[
    e_n=(K^3/n)^{1/2},
    \qquad
    \widehat{\bs T}_K
    =\bs G_K^{-1/2}\widehat{\bs G}_K\bs G_K^{-1/2}.
\]
The conditions \(K^5/n\to0\) and
\(n/N=\rho_n/(1-\rho_n)=O(1)\) imply
\(e_n=o(1)\) and \(K/\sqrt m=o(1)\) for \(m\in\{n,N\}\).
Since
\[
    \mathbb E_m\{(\mathcal C_m\bs Z_K)^\top
                  \mathcal C_m\bs Z_K\}
    =
    \mathbb E_m(\bs Z_K^\top\bs Z_K)
    -(\mathbb E_m\bs Z_K)^\top\mathbb E_m\bs Z_K,
\]
\eqref{supp:eq:fixed-design-moments} gives
\begin{align*}
    &\|\mathbb E_m\{(\mathcal C_m\bs Z_K)^\top
                    \mathcal C_m\bs Z_K\}-\bs I_{K+1}\|_{\rm op}\\
    &\qquad\leq
    \|\mathbb E_m(\bs Z_K^\top\bs Z_K)-\bs I_{K+1}\|_{\rm op}
    +\|\mathbb E_m\bs Z_K\|_F^2\\
    &\qquad=O_{\mathbb P}(K/\sqrt m+K/m)=o_{\mathbb P}(1).
\end{align*}

For matrix-valued functions \(H_1,H_2\) evaluated on the same sample,
Cauchy--Schwarz gives
\begin{equation}
\begin{aligned}
    \|\mathbb E_m(H_1^\top H_2)\|_{\rm op}
    &=
    \sup_{\|\bs u\|=\|\bs v\|=1}
    |\mathbb E_m\{(H_1\bs u)^\top H_2\bs v\}|\\
    &\leq
    \|\mathbb E_m(H_1^\top H_1)\|_{\rm op}^{1/2}
    \|\mathbb E_m(H_2^\top H_2)\|_{\rm op}^{1/2}.
\end{aligned}
\label{supp:eq:operator-cauchy-schwarz}
\end{equation}
Put
\[
    \bs A_K=\mathcal C_n\bs Z_K,
    \qquad
    \bs E_K=\mathcal C_n(\widehat{\bs Z}_K-\bs Z_K).
\]
By \eqref{supp:eq:design-perturbation},
\[
    \mathbb E_n\|\bs E_K\|_F^2
    \leq
    \mathbb E_n\|\widehat{\bs Z}_K-\bs Z_K\|_F^2
    =O_{\mathbb P}(e_n^2).
\]
Consequently,
\begin{align*}
    \|\mathbb E_n(\bs A_K^\top\bs E_K)\|_{\rm op}
    &\leq
    \|\mathbb E_n(\bs A_K^\top\bs A_K)\|_{\rm op}^{1/2}
    \{\mathbb E_n\|\bs E_K\|_F^2\}^{1/2}
    =O_{\mathbb P}(e_n),\\
    \|\mathbb E_n(\bs E_K^\top\bs E_K)\|_{\rm op}
    &\leq\mathbb E_n\|\bs E_K\|_F^2
    =O_{\mathbb P}(e_n^2).
\end{align*}
Using \(\mathcal C_n\widehat{\bs Z}_K=\bs A_K+\bs E_K\),
\begin{align*}
    &\mathbb E_n\{(\mathcal C_n\widehat{\bs Z}_K)^\top
                    \mathcal C_n\widehat{\bs Z}_K\}
      -\mathbb E_n(\bs A_K^\top\bs A_K)\\
    &\qquad=
    \mathbb E_n(\bs A_K^\top\bs E_K)
    +\mathbb E_n(\bs E_K^\top\bs A_K)
    +\mathbb E_n(\bs E_K^\top\bs E_K).
\end{align*}
It follows that
\[
    \|\mathbb E_n\{(\mathcal C_n\widehat{\bs Z}_K)^\top
                    \mathcal C_n\widehat{\bs Z}_K\}\|_{\rm op}
    =O_{\mathbb P}(1).
\]

By \eqref{supp:eq:target-rates},
\(\|\widehat{\bs J}-\bs J\|_{\rm op}
=O_{\mathbb P}(n^{-1/2})\).
Since \(\bs J\) is nonsingular,
\[
    \|\widehat{\bs J}^{-1}\|_{\rm op}=O_{\mathbb P}(1),
    \qquad
    \|\widehat{\bs J}^{-1}\bs J-\bs I_p\|_{\rm op}
    \leq
    \|\widehat{\bs J}^{-1}\|_{\rm op}
    \|\bs J-\widehat{\bs J}\|_{\rm op}
    =O_{\mathbb P}(n^{-1/2}).
\]
For
\[
    \widetilde{\bs M}_n
    =\bs J^\top\widehat{\bs M}\bs J
    =(\widehat{\bs J}^{-1}\bs J)^\top
      (\widehat{\bs J}^{-1}\bs J),
\]
we therefore have
\begin{align*}
    \widetilde{\bs M}_n-\bs I_p
    &=
    (\widehat{\bs J}^{-1}\bs J-\bs I_p)^\top
    +(\widehat{\bs J}^{-1}\bs J-\bs I_p)\\
    &\quad+
    (\widehat{\bs J}^{-1}\bs J-\bs I_p)^\top
    (\widehat{\bs J}^{-1}\bs J-\bs I_p).
\end{align*}
Hence
\begin{align*}
    \|\widetilde{\bs M}_n-\bs I_p\|_{\rm op}
    &\leq
    2\|\widehat{\bs J}^{-1}\bs J-\bs I_p\|_{\rm op}
    +\|\widehat{\bs J}^{-1}\bs J-\bs I_p\|_{\rm op}^2\\
    &=O_{\mathbb P}(n^{-1/2})+O_{\mathbb P}(n^{-1})
    =O_{\mathbb P}(n^{-1/2}).
\end{align*}

The definition of \(\widehat{\bs G}_K\) gives
\[
    \widehat{\bs T}_K
    =
    \mathbb E_n\{(\mathcal C_n\widehat{\bs Z}_K)^\top
        \widetilde{\bs M}_n
        \mathcal C_n\widehat{\bs Z}_K\}.
\]
Hence
\begin{align*}
    \widehat{\bs T}_K-\bs I_{K+1}
    &=
    \mathbb E_n(\bs Z_K^\top\bs Z_K)-\bs I_{K+1}
    -(\mathbb E_n\bs Z_K)^\top\mathbb E_n\bs Z_K\\
    &\quad+
    \mathbb E_n(\bs A_K^\top\bs E_K)
    +\mathbb E_n(\bs E_K^\top\bs A_K)
    +\mathbb E_n(\bs E_K^\top\bs E_K)\\
    &\quad+
    \mathbb E_n\{(\mathcal C_n\widehat{\bs Z}_K)^\top
        (\widetilde{\bs M}_n-\bs I_p)
        \mathcal C_n\widehat{\bs Z}_K\}.
\end{align*}
The last term satisfies
\begin{align*}
    &\|\mathbb E_n\{(\mathcal C_n\widehat{\bs Z}_K)^\top
        (\widetilde{\bs M}_n-\bs I_p)
        \mathcal C_n\widehat{\bs Z}_K\}\|_{\rm op}\\
    &\qquad\leq
    \|\widetilde{\bs M}_n-\bs I_p\|_{\rm op}
    \|\mathbb E_n\{(\mathcal C_n\widehat{\bs Z}_K)^\top
                    \mathcal C_n\widehat{\bs Z}_K\}\|_{\rm op}\\
    &\qquad=O_{\mathbb P}(n^{-1/2}).
\end{align*}
Combining these bounds gives
\[
    \|\widehat{\bs T}_K-\bs I_{K+1}\|_{\rm op}
    =
    O_{\mathbb P}(K/\sqrt n+K/n+e_n+e_n^2+n^{-1/2})
    =O_{\mathbb P}(e_n).
\]
Define
\[
    \mathcal G_n
    =\{\|\widehat{\bs T}_K-\bs I_{K+1}\|_{\rm op}\leq1/2\}.
\]
Then \(\mathbb P(\mathcal G_n)\to1\), and on \(\mathcal G_n\),
\[
    \lambda_{\min}(\widehat{\bs T}_K)\geq\tfrac12,
    \qquad
    \|\widehat{\bs T}_K^{-1}\|_{\rm op}\leq2,
    \qquad
    \widehat{\bs G}_K
    =\bs G_K^{1/2}\widehat{\bs T}_K\bs G_K^{1/2}
    \succ0.
\]

\smallskip\noindent
\textit{Step 4: Bounding
\(\left\|\bs G_K^{1/2}
\left(\widehat{\bs\beta}_{\mathrm{san},K}
-\frac{1-\rho_n}{1-\rho}\bs\beta_{\mathrm{san},K}\right)\right\|\).}
Write
\[
    \bs a_K
    =\E(\bs Z_K^\top\bs J^{-1}\bs\psi^\ast)
    =\bs G_K^{-1/2}\bs d_K.
\]
By \eqref{supp:eq:normalized-design},
\begin{align*}
    0
    &\leq
    \E\|\bs J^{-1}\bs\psi^\ast-\bs Z_K\bs a_K\|^2\\
    &=
    \E\|\bs J^{-1}\bs\psi^\ast\|^2
    -2\bs a_K^\top
        \E(\bs Z_K^\top\bs J^{-1}\bs\psi^\ast)
    +\bs a_K^\top\E(\bs Z_K^\top\bs Z_K)\bs a_K\\
    &=
    \E\|\bs J^{-1}\bs\psi^\ast\|^2-\|\bs a_K\|^2.
\end{align*}
Thus
\begin{equation}
    \|\bs a_K\|^2
    \leq\E\|\bs J^{-1}\bs\psi^\ast\|^2=O(1).
    \label{supp:eq:projection-bound}
\end{equation}

Assumption~\ref{ass:score-learning}(ii) gives
\[
    \E\|\bs J^{-1}\bs\psi^\ast\|^4
    \leq
    C\E\{\|\widetilde{\bs X}\|^4(|Y|^4+D^4)\}
    <\infty.
\]
Together with \eqref{supp:eq:design-envelopes}, this implies
\begin{align*}
    &\E\|\mathbb E_n(\bs Z_K^\top\bs J^{-1}\bs\psi^\ast)
             -\bs a_K\|^2\\
    &\qquad\leq
    \frac1n\E(\|\bs Z_K\|_F^2
                    \|\bs J^{-1}\bs\psi^\ast\|^2)\\
    &\qquad\leq
    \frac{CK}{n}
    (\E F^4)^{1/2}
    (\E\|\bs J^{-1}\bs\psi^\ast\|^4)^{1/2}
    =O(K/n).
\end{align*}
Also, \(\E\bs\psi^\ast=\bs0\) implies
\[
    \E\|\bs J^{-1}\mathbb E_n\bs\psi^\ast\|^2
    =\frac1n\E\|\bs J^{-1}\bs\psi^\ast\|^2
    =O(n^{-1}).
\]
Consequently,
\begin{align*}
    \|\mathbb E_n(\bs Z_K^\top\bs J^{-1}\bs\psi^\ast)
             -\bs a_K\|
    &=O_{\mathbb P}(\sqrt{K/n}),\\
    \|(\mathbb E_n\bs Z_K)^\top
             \bs J^{-1}\mathbb E_n\bs\psi^\ast\|
    &\leq
    \|\mathbb E_n\bs Z_K\|_F
    \|\bs J^{-1}\mathbb E_n\bs\psi^\ast\|\\
    &=O_{\mathbb P}(\sqrt K/n).
\end{align*}

Since \(\mathbb E_n\mathcal C_n\widehat{\bs Z}_K=\bs0\),
the definition of \(\widehat{\bs d}_K\) yields
\[
    \bs G_K^{-1/2}\widehat{\bs d}_K
    =
    \mathbb E_n\{(\mathcal C_n\widehat{\bs Z}_K)^\top
        \widetilde{\bs M}_n\bs J^{-1}\widehat{\bs\psi}\}.
\]
Using
\(\mathcal C_n\widehat{\bs Z}_K=\mathcal C_n\bs Z_K+\bs E_K\),
we obtain
\begin{align*}
    \bs G_K^{-1/2}\widehat{\bs d}_K-\bs a_K
    &=
    \mathbb E_n(\bs Z_K^\top\bs J^{-1}\bs\psi^\ast)
    -\bs a_K
    -(\mathbb E_n\bs Z_K)^\top
       \bs J^{-1}\mathbb E_n\bs\psi^\ast\\
    &\quad+
    \mathbb E_n(\bs E_K^\top\bs J^{-1}\bs\psi^\ast)\\
    &\quad+
    \mathbb E_n\{(\mathcal C_n\widehat{\bs Z}_K)^\top
        (\widetilde{\bs M}_n-\bs I_p)\bs J^{-1}\bs\psi^\ast\}\\
    &\quad+
    \mathbb E_n\{(\mathcal C_n\widehat{\bs Z}_K)^\top
        \widetilde{\bs M}_n\bs J^{-1}
        (\widehat{\bs\psi}-\bs\psi^\ast)\}.
\end{align*}
The last three terms satisfy
\begin{align*}
    &\|\mathbb E_n(\bs E_K^\top\bs J^{-1}\bs\psi^\ast)\|\\
    &\qquad\leq
    \{\mathbb E_n\|\bs E_K\|_F^2\}^{1/2}
    \{\mathbb E_n\|\bs J^{-1}\bs\psi^\ast\|^2\}^{1/2}
    =O_{\mathbb P}(e_n),\\
    &\|\mathbb E_n\{(\mathcal C_n\widehat{\bs Z}_K)^\top
        (\widetilde{\bs M}_n-\bs I_p)
        \bs J^{-1}\bs\psi^\ast\}\|\\
    &\qquad\leq
    \|\mathbb E_n\{(\mathcal C_n\widehat{\bs Z}_K)^\top
                    \mathcal C_n\widehat{\bs Z}_K\}\|_{\rm op}^{1/2}
    \|\widetilde{\bs M}_n-\bs I_p\|_{\rm op}
    \{\mathbb E_n\|\bs J^{-1}\bs\psi^\ast\|^2\}^{1/2}\\
    &\qquad=O_{\mathbb P}(n^{-1/2}),\\
    &\|\mathbb E_n\{(\mathcal C_n\widehat{\bs Z}_K)^\top
        \widetilde{\bs M}_n\bs J^{-1}
        (\widehat{\bs\psi}-\bs\psi^\ast)\}\|\\
    &\qquad\leq
    \|\mathbb E_n\{(\mathcal C_n\widehat{\bs Z}_K)^\top
                    \mathcal C_n\widehat{\bs Z}_K\}\|_{\rm op}^{1/2}
    \|\widetilde{\bs M}_n\|_{\rm op}
    \|\bs J^{-1}\|_{\rm op}
    \{\mathbb E_n\|\widehat{\bs\psi}-\bs\psi^\ast\|^2\}^{1/2}\\
    &\qquad=O_{\mathbb P}(n^{-1/2}),
\end{align*}
where the last bound uses \eqref{supp:eq:target-rates}.
Therefore,
\[
    \|\bs G_K^{-1/2}\widehat{\bs d}_K-\bs a_K\|
    =
    O_{\mathbb P}(\sqrt{K/n}+\sqrt K/n+e_n+n^{-1/2})
    =O_{\mathbb P}(e_n).
\]

On \(\mathcal G_n\), define
\[
    \widehat{\bs a}_K
    =\frac{\bs G_K^{1/2}\widehat{\bs\beta}_{\mathrm{san},K}}
           {1-\rho_n}.
\]
Since
\[
    \widehat{\bs G}_K\widehat{\bs\beta}_{\mathrm{san},K}
    =(1-\rho_n)\widehat{\bs d}_K,
\]
the definitions of \(\widehat{\bs T}_K\) and
\(\widehat{\bs a}_K\) give
\[
\begin{aligned}
    \widehat{\bs T}_K\widehat{\bs a}_K
    &=
    (\bs G_K^{-1/2}\widehat{\bs G}_K\bs G_K^{-1/2})
    \frac{\bs G_K^{1/2}\widehat{\bs\beta}_{\mathrm{san},K}}
         {1-\rho_n}\\
    &=
    \frac{\bs G_K^{-1/2}
          \widehat{\bs G}_K\widehat{\bs\beta}_{\mathrm{san},K}}
         {1-\rho_n}
    =\bs G_K^{-1/2}\widehat{\bs d}_K.
\end{aligned}
\]
Hence
\[
    \widehat{\bs a}_K-\bs a_K
    =
    \widehat{\bs T}_K^{-1}
    \{\bs G_K^{-1/2}\widehat{\bs d}_K-\bs a_K
      -(\widehat{\bs T}_K-\bs I_{K+1})\bs a_K\}.
\]
Using \eqref{supp:eq:projection-bound},
\begin{equation}
\begin{aligned}
    \|\widehat{\bs a}_K-\bs a_K\|
    &\leq
    2\{\|\bs G_K^{-1/2}\widehat{\bs d}_K-\bs a_K\|
      +\|\widehat{\bs T}_K-\bs I_{K+1}\|_{\rm op}\|\bs a_K\|\}\\
    &=O_{\mathbb P}(e_n).
\end{aligned}
\label{supp:eq:normalized-coefficient}
\end{equation}
Finally,
\(\bs G_K^{1/2}\bs\beta_{\mathrm{san},K}
=(1-\rho)\bs a_K\), so
\[
    \left\|
    \bs G_K^{1/2}
    \left(
        \widehat{\bs\beta}_{\mathrm{san},K}
        -\frac{1-\rho_n}{1-\rho}\bs\beta_{\mathrm{san},K}
    \right)
    \right\|
    =
    (1-\rho_n)\|\widehat{\bs a}_K-\bs a_K\|
    =
    O_{\mathbb P}\{(K^3/n)^{1/2}\}.
\]
\end{proof}

\begin{proof}[Proof of Lemma~\ref{supp:lem:score-replacement}]
Use \(\bs Z_{\bs\eta,K}\), \(\bs Z_K\),
\(\widehat{\bs Z}_K\), \(\bs a_K\), and
\(\widehat{\bs a}_K\) from the proof of
Lemma~\ref{supp:lem:normal-equations}, and write
\[
    e_n=(K^3/n)^{1/2},
    \qquad
    \bs t_n=\widehat{\bs\eta}-\bs\eta^\ast.
\]
By \eqref{supp:eq:projection-bound},
\eqref{supp:eq:normalized-coefficient}, and
\(\|\bs t_n\|=O_{\mathbb P}(n^{-1/2})\),
\[
    \|\bs a_K\|=O(1),
    \qquad
    \|\widehat{\bs a}_K-\bs a_K\|=O_{\mathbb P}(e_n),
    \qquad
    \|\widehat{\bs a}_K\|
    \leq\|\bs a_K\|+\|\widehat{\bs a}_K-\bs a_K\|
    =O_{\mathbb P}(1).
\]

\smallskip\noindent
\textit{Step 1: Bounding \(\sqrt n\|\Delta_n\bs\delta_K\|\).}

Write
\[
    \dot{\bs Z}_K
    =
    \left.\partial_{\bs\eta}\bs Z_{\bs\eta,K}
    \right|_{\bs\eta=\bs\eta^\ast}.
\]
By \eqref{supp:eq:two-sample-moment},
\eqref{supp:eq:normalized-design}, and
\eqref{supp:eq:design-envelopes},
\begin{align*}
    \E\|\sqrt n\Delta_n\bs Z_K\|_F^2
    &=(1+n/N)\E\|\bs Z_K\|_F^2
    \leq C(K+1),\\
    \E\|\sqrt n\Delta_n\dot{\bs Z}_K\|_F^2
    &=(1+n/N)\E\|\mathcal C\dot{\bs Z}_K\|_F^2\\
    &\leq(1+n/N)\E\|\dot{\bs Z}_K\|_F^2
    \leq CK^3\E F^2.
\end{align*}
Thus
\[
    \sqrt n\|\Delta_n\bs Z_K\|_F
    =O_{\mathbb P}(K^{1/2}),
    \qquad
    \sqrt n\|\Delta_n\dot{\bs Z}_K\|_F
    =O_{\mathbb P}(K^{3/2}).
\]

Choose a fixed \(\delta>0\) such that
\[
    \{\bs\eta:\|\bs\eta-\bs\eta^\ast\|\leq\delta\}
    \subseteq\mathcal U_\theta\times\mathcal U_\gamma.
\]
For \(\bs\eta=(\bs\theta^\top,\bs\gamma^\top)^\top\)
and \(\bs\eta'=(\bs\theta'^\top,\bs\gamma'^\top)^\top\)
in this ball, Assumption~\ref{ass:score-learning}(ii)
gives
\begin{align*}
    |S_{\bs\gamma}-S_{\bs\gamma'}|
    &\leq
    \int_0^1
    \|\partial_{\bs\gamma}
      S_{\bs\gamma'+u(\bs\gamma-\bs\gamma')}\|\,du\,
    \|\bs\gamma-\bs\gamma'\|\\
    &\leq D^{1/2}\|\bs\gamma-\bs\gamma'\|,\\
    \|\partial_{\bs\gamma}S_{\bs\gamma}
      -\partial_{\bs\gamma}S_{\bs\gamma'}\|
    &\leq
    \int_0^1
    \|\partial_{\bs\gamma}^2
      S_{\bs\gamma'+u(\bs\gamma-\bs\gamma')}\|_{\rm op}
    \,du\,\|\bs\gamma-\bs\gamma'\|\\
    &\leq D\|\bs\gamma-\bs\gamma'\|,\\
    \|\partial_{\bs\theta}\mu_{\bs\theta}
      -\partial_{\bs\theta}\mu_{\bs\theta'}\|
    &\leq
    \int_0^1
    \|\partial_{\bs\theta}^2
      \mu_{\bs\theta'+u(\bs\theta-\bs\theta')}\|_{\rm op}
    \,du\,\|\bs\theta-\bs\theta'\|\\
    &\leq D\|\bs\theta-\bs\theta'\|.
\end{align*}
Since
\begin{align*}
    &\partial_{\bs\gamma}\bs b_K(S_{\bs\gamma})
     -\partial_{\bs\gamma}\bs b_K(S_{\bs\gamma'})\\
    &\quad=
    \{\bs b_K'(S_{\bs\gamma})-\bs b_K'(S_{\bs\gamma'})\}
    (\partial_{\bs\gamma}S_{\bs\gamma})^\top\\
    &\qquad+
    \bs b_K'(S_{\bs\gamma'})
    (\partial_{\bs\gamma}S_{\bs\gamma}
     -\partial_{\bs\gamma}S_{\bs\gamma'})^\top,
\end{align*}
Lemma~\ref{supp:lem:spline-bounds} implies
\begin{align*}
    &\|\partial_{\bs\gamma}\bs b_K(S_{\bs\gamma})
      -\partial_{\bs\gamma}\bs b_K(S_{\bs\gamma'})\|_F\\
    &\quad\leq
    \|\bs b_K'(S_{\bs\gamma})-\bs b_K'(S_{\bs\gamma'})\|
    \|\partial_{\bs\gamma}S_{\bs\gamma}\|\\
    &\qquad+
    \|\bs b_K'(S_{\bs\gamma'})\|
    \|\partial_{\bs\gamma}S_{\bs\gamma}
      -\partial_{\bs\gamma}S_{\bs\gamma'}\|\\
    &\quad\leq
    CK^2|S_{\bs\gamma}-S_{\bs\gamma'}|D^{1/2}
    +CKD\|\bs\gamma-\bs\gamma'\|\\
    &\quad\leq CK^2D\|\bs\gamma-\bs\gamma'\|.
\end{align*}
Therefore,
\begin{align*}
    &\|\partial_{\bs\eta}\bs B_{\bs\eta,K}
      -\partial_{\bs\eta}\bs B_{\bs\eta',K}\|_F\\
    &\quad=
    \|\widetilde{\bs X}\|
    \bigl\{
        \|\partial_{\bs\theta}\mu_{\bs\theta}
          -\partial_{\bs\theta}\mu_{\bs\theta'}\|^2
        +\|\partial_{\bs\gamma}\bs b_K(S_{\bs\gamma})
          -\partial_{\bs\gamma}\bs b_K(S_{\bs\gamma'})\|_F^2
    \bigr\}^{1/2}\\
    &\quad\leq
    \|\widetilde{\bs X}\|D
    \{\|\bs\theta-\bs\theta'\|^2
      +CK^4\|\bs\gamma-\bs\gamma'\|^2\}^{1/2}\\
    &\quad\leq
    CK^2\|\widetilde{\bs X}\|D\|\bs\eta-\bs\eta'\|.
\end{align*}
Using
\(F=\|\widetilde{\bs X}\|D+\E(\|\widetilde{\bs X}\|D)\)
and \(\|\bs G_K^{-1/2}\|_{\rm op}\leq CK^{1/2}\),
we obtain
\begin{equation}
\begin{aligned}
    &\|\partial_{\bs\eta}\bs Z_{\bs\eta,K}
      -\partial_{\bs\eta}\bs Z_{\bs\eta',K}\|_F\\
    &\quad\leq
    \|\bs J^{-1}\|_{\rm op}\|\bs G_K^{-1/2}\|_{\rm op}
    \bigl[
        \|\partial_{\bs\eta}\bs B_{\bs\eta,K}
          -\partial_{\bs\eta}\bs B_{\bs\eta',K}\|_F\\
    &\hspace{48mm}
        +\E\|\partial_{\bs\eta}\bs B_{\bs\eta,K}
             -\partial_{\bs\eta}\bs B_{\bs\eta',K}\|_F
    \bigr]\\
    &\quad\leq CK^{5/2}F\|\bs\eta-\bs\eta'\|.
\end{aligned}
\label{supp:eq:design-lipschitz}
\end{equation}

Since \(\|\bs t_n\|=O_{\mathbb P}(n^{-1/2})\),
\(\mathbb P(\|\bs t_n\|\leq\delta)\to1\).
On this event,
\(\|\bs\eta^\ast+u\bs t_n-\bs\eta^\ast\|\leq\delta\)
for every \(u\in[0,1]\), and
\begin{align*}
    \widehat{\bs Z}_K-\bs Z_K
    &=\dot{\bs Z}_K\bs t_n+\bs R_K,\\
    \bs R_K
    &=
    \int_0^1
    \{\partial_{\bs\eta}\bs Z_{\bs\eta^\ast+u\bs t_n,K}
      -\dot{\bs Z}_K\}\bs t_n\,du.
\end{align*}
By \eqref{supp:eq:design-lipschitz},
\[
    \|\bs R_K\|_F
    \leq
    \int_0^1 CK^{5/2}Fu\|\bs t_n\|^2\,du
    \leq CK^{5/2}F\|\bs t_n\|^2.
\]
Consequently,
\begin{align*}
    \sqrt n\Delta_n(\widehat{\bs Z}_K-\bs Z_K)
    &=(\sqrt n\Delta_n\dot{\bs Z}_K)\bs t_n
      +\sqrt n\Delta_n\bs R_K,\\
    \sqrt n\|\Delta_n(\widehat{\bs Z}_K-\bs Z_K)\|_F
    &\leq
    \sqrt n\|\Delta_n\dot{\bs Z}_K\|_F\|\bs t_n\|
    +\sqrt n(\mathbb E_n+\mathbb E_N)\|\bs R_K\|_F\\
    &\leq
    \sqrt n\|\Delta_n\dot{\bs Z}_K\|_F\|\bs t_n\|
    +C\sqrt nK^{5/2}\|\bs t_n\|^2
       (\mathbb E_nF+\mathbb E_NF).
\end{align*}
Because \(\E F<\infty\),
\(\mathbb E_nF+\mathbb E_NF=O_{\mathbb P}(1)\), and hence
\begin{equation}
\begin{aligned}
    \sqrt n\|\Delta_n(\widehat{\bs Z}_K-\bs Z_K)\|_F
    &=
    O_{\mathbb P}(K^{3/2})O_{\mathbb P}(n^{-1/2})
    +\sqrt nK^{5/2}O_{\mathbb P}(n^{-1})
       O_{\mathbb P}(1)\\
    &=O_{\mathbb P}(K^{5/2}/\sqrt n).
\end{aligned}
\label{supp:eq:taylor-replacement}
\end{equation}

Recall that
\[
    \bs\delta_K
    =\bs B_{\widehat{\bs\eta},K}
       \widehat{\bs\beta}_{\mathrm{san},K}
     -\frac{1-\rho_n}{1-\rho}
       \bs B_{\bs\eta^\ast,K}\bs\beta_{\mathrm{san},K}.
\]
The definitions of \(\bs a_K\) and \(\widehat{\bs a}_K\) give
\begin{equation}
    \bs\delta_K
    =(1-\rho_n)
    \{\bs B_{\widehat{\bs\eta},K}\bs G_K^{-1/2}
         \widehat{\bs a}_K
      -\bs B_{\bs\eta^\ast,K}\bs G_K^{-1/2}\bs a_K\}.
    \label{supp:eq:score-difference-decomposition}
\end{equation}
For each \(\bs\eta\),
\[
    \bs B_{\bs\eta,K}\bs G_K^{-1/2}
    =
    \bs J\bs Z_{\bs\eta,K}
    +(\E\bs B_{\bs\eta,K})\bs G_K^{-1/2}.
\]
The second term is constant across observations and is therefore
annihilated by \(\Delta_n\). Thus
\begin{align*}
    \sqrt n\Delta_n\bs\delta_K
    &=(1-\rho_n)\bs J
      \{(\sqrt n\Delta_n\widehat{\bs Z}_K)\widehat{\bs a}_K
        -(\sqrt n\Delta_n\bs Z_K)\bs a_K\}\\
    &=(1-\rho_n)\bs J
      \{(\sqrt n\Delta_n\bs Z_K)
          (\widehat{\bs a}_K-\bs a_K)
        +\sqrt n\Delta_n(\widehat{\bs Z}_K-\bs Z_K)
          \widehat{\bs a}_K\}.
\end{align*}
It follows that
\begin{align*}
    \sqrt n\|\Delta_n\bs\delta_K\|
    &\leq C\{
        \sqrt n\|\Delta_n\bs Z_K\|_F
            \|\widehat{\bs a}_K-\bs a_K\|\\
    &\hspace{19mm}
        +\sqrt n\|\Delta_n(\widehat{\bs Z}_K-\bs Z_K)\|_F
            \|\widehat{\bs a}_K\|
      \}\\
    &=O_{\mathbb P}(K^{1/2})
        O_{\mathbb P}(K^{3/2}/\sqrt n)
      +O_{\mathbb P}(K^{5/2}/\sqrt n)O_{\mathbb P}(1)\\
    &=O_{\mathbb P}\{(K^5/n)^{1/2}\}
    =o_{\mathbb P}(1),
\end{align*}
which proves \eqref{supp:eq:mean-replacement}.

\smallskip\noindent
\textit{Step 2: Bounding
\(\tr\{\widehat{\bs V}_m(\bs\delta_K)\}\)
for \(m\in\{n,N\}\).}

Applying \(\mathcal C_m\) to
\eqref{supp:eq:score-difference-decomposition} gives
\begin{align*}
    \mathcal C_m\bs\delta_K
    &=(1-\rho_n)\bs J
      \{(\mathcal C_m\widehat{\bs Z}_K)\widehat{\bs a}_K
        -(\mathcal C_m\bs Z_K)\bs a_K\}\\
    &=(1-\rho_n)\bs J
      \{(\mathcal C_m\bs Z_K)
           (\widehat{\bs a}_K-\bs a_K)
        +\mathcal C_m(\widehat{\bs Z}_K-\bs Z_K)
           \widehat{\bs a}_K\}.
\end{align*}
For the first term,
\begin{align*}
    &\mathbb E_m
    \|(\mathcal C_m\bs Z_K)(\widehat{\bs a}_K-\bs a_K)\|^2\\
    &\quad=
    (\widehat{\bs a}_K-\bs a_K)^\top
    \mathbb E_m\{(\mathcal C_m\bs Z_K)^\top
                   \mathcal C_m\bs Z_K\}
    (\widehat{\bs a}_K-\bs a_K)\\
    &\quad\leq
    \|\mathbb E_m(\bs Z_K^\top\bs Z_K)\|_{\rm op}
    \|\widehat{\bs a}_K-\bs a_K\|^2\\
    &\quad=O_{\mathbb P}(1)O_{\mathbb P}(e_n^2)
    =O_{\mathbb P}(e_n^2),
\end{align*}
where \eqref{supp:eq:fixed-design-moments} and
\(K/\sqrt m=o(1)\) give
\(\|\mathbb E_m(\bs Z_K^\top\bs Z_K)\|_{\rm op}
=O_{\mathbb P}(1)\).
For the second term, \eqref{supp:eq:design-perturbation} gives
\begin{align*}
    &\mathbb E_m
    \|\mathcal C_m(\widehat{\bs Z}_K-\bs Z_K)
       \widehat{\bs a}_K\|^2\\
    &\quad\leq
    \mathbb E_m
    \|\mathcal C_m(\widehat{\bs Z}_K-\bs Z_K)\|_F^2
    \|\widehat{\bs a}_K\|^2\\
    &\quad\leq
    \mathbb E_m\|\widehat{\bs Z}_K-\bs Z_K\|_F^2
    \|\widehat{\bs a}_K\|^2\\
    &\quad=O_{\mathbb P}(e_n^2)O_{\mathbb P}(1)
    =O_{\mathbb P}(e_n^2).
\end{align*}
Therefore,
\begin{align*}
    \tr\{\widehat{\bs V}_m(\bs\delta_K)\}
    &=
    \mathbb E_m\tr\{
        (\mathcal C_m\bs\delta_K)
        (\mathcal C_m\bs\delta_K)^\top
    \}\\
    &=\mathbb E_m\|\mathcal C_m\bs\delta_K\|^2\\
    &\leq
    2(1-\rho_n)^2\|\bs J\|_{\rm op}^2
    \bigl\{
        \mathbb E_m
        \|(\mathcal C_m\bs Z_K)
          (\widehat{\bs a}_K-\bs a_K)\|^2\\
    &\hspace{43mm}
        +\mathbb E_m
        \|\mathcal C_m(\widehat{\bs Z}_K-\bs Z_K)
          \widehat{\bs a}_K\|^2
    \bigr\}\\
    &=O_{\mathbb P}(e_n^2)
    =O_{\mathbb P}(K^3/n)
    =o_{\mathbb P}(1).
\end{align*}
Summing over \(m=n,N\) proves
\eqref{supp:eq:variance-replacement}.
\end{proof}

\end{document}